\documentclass[12pt]{article}

\usepackage{caption,subcaption}

\usepackage{appendix}

\usepackage{amsfonts,amsmath,amssymb,accents,mathrsfs}
\usepackage[retainorgcmds]{IEEEtrantools}
\usepackage{multirow}
\usepackage{multicol}
\usepackage{tikz}
\usepackage{graphicx,color}
\usepackage{booktabs}
\usepackage{placeins}
\usepackage[colorlinks,linkcolor=Blue,citecolor=Blue,urlcolor=Blue,bookmarks,
bookmarksnumbered]{hyperref}
\usepackage[nosort]{cite}
\usepackage[scaled=0.85]{helvet}
\usepackage[T1]{fontenc}
\usepackage[utf8]{inputenc}
\usepackage{cancel}

\definecolor{Red}    {rgb}{0.90,0.00,0.12} 
\definecolor{Blue}   {rgb}{0.00,0.00,1.00} 
\definecolor{Green}  {rgb}{0.10,0.70,0.10} 
\definecolor{Turque} {rgb}{0.00,0.65,0.85} 
\definecolor{Orange} {rgb}{1.00,0.50,0.15} 
\definecolor{Magenta}{rgb}{1.00,0.00,1.00} 
\definecolor{Gold}   {rgb}{1.00,0.75,0.25} 
\definecolor{Seaweed}{rgb}{0.01,0.24,0.09} 
\definecolor{Purple} {rgb}{0.50,0.25,0.55} 
\definecolor{Brown}  {rgb}{0.43,0.26,0.32} 
\definecolor{grey1}  {rgb}{0.20,0.20,0.20} 
\definecolor{grey2}  {rgb}{0.40,0.40,0.40} 
\definecolor{grey3}  {rgb}{0.60,0.60,0.60} 
\definecolor{grey4}  {rgb}{0.80,0.80,0.80} 
\definecolor{grey5}  {rgb}{0.90,0.90,0.90} 

\def\a{{\alpha}}
\def\b{{\beta}}

\def\ad{{\dot{\alpha}}}
\def\bd{{\dot{\beta}}}

\def\N{{\mathcal{N}}}

\def\Ysf{{\textsf{Y}}}

\def\D{{\rm D}}
\def\Dd{{\bar{\rm D}}}
\def\pa{\partial}

\def\ff#1-#2{\frac{#1}{#2}}
\def\tff#1-#2{\tfrac{#1}{#2}}

\def\be{\begin{equation}}
\def\ee{\end{equation}}
\def\beq{\begin{equation}}
\def\eeq{\end{equation}}
\def\bIea{\begin{IEEEeqnarray*}}
\def\eIea{\end{IEEEeqnarray*}}
\def\bea{\begin{eqnarray}}
\def\eea{\end{eqnarray}}
\def\n{\IEEEyesnumber}
\def\sn{\IEEEyessubnumber}

\makeatletter
\def\section{\@startsection{section}{1}{\z@}
              {3ex plus-1ex minus-.2ex}{1pt plus1pt}
              {\large\sf\bfseries\boldmath}}
\def\subsection{\@startsection{subsection}{2}{\z@}
              {1.5ex plus-1ex minus-.2ex}{0.01pt plus1pt}{\sf\slshape}}
\def\subsubsection{\@startsection{subsubsection}{3}{\z@}
              {1.5ex plus-1ex minus-.2ex}{0.01pt plus0.2pt}{\sf\boldmath}}
\def\paragraph{\@startsection{paragraph}{4}{\z@}
              {.75ex \@plus.5ex \@minus.2ex}{-2mm}{\sf\bfseries\boldmath}}
\makeatother

   \parskip\medskipamount     \lineskip=0pt
\thicklines
\begin{document}
\thispagestyle{empty}
\noindent{\small
\vspace*{6mm}
\begin{center}
{\large \bf 
Towards a Direct Detection of the Spin of Dark Matter
\vspace{3ex}
} \\   [9mm] {\large { 
Leah Jenks\footnote{ljenks@uchicago.edu}$^{a}$,
K.\ Koutrolikos\footnote{koutrol@umd.edu}$^{b}$
 Evan McDonough\footnote{e.mcdonough@uwinnipeg.ca}$^{c}$,
Stephon Alexander\footnote{stephon\_alexander@brown.edu}$^{d}$,\\
and S.\ James Gates Jr.\footnote{gatess@umd.edu}$^{b}$
}
}
\\*
 \emph{
 \centering
  \\[6pt]
 $^{a}$  Kavli Institute for Cosmological Physics,
 \\[1pt]
  University of Chicago, Chicago, IL 60637, USA\\[6pt]
  $^{b}$ Department of Physics, University of Maryland,\\[1pt]
College Park, MD 20742-4111, USA
 \\[6pt]
   $^{c}$ {Department of Physics, University of Winnipeg, \\[1pt] Winnipeg, MB R3B 2E9 Canada}
 \\[6pt]
 $^{d}$Brown Theoretical Physics Center and Department of Physics,\\[1pt] Brown University, Providence, RI 02912, USA
 }
  
$$~~$$
  $$~~$$
 \\*[-8mm]
{ ABSTRACT}\\[4mm]
\parbox{142mm}{\parindent=2pc\indent\baselineskip=14pt plus1pt
We investigate the contribution of higher spin particles in the signal of direct detection searches for dark matter. We consider a bosonic or fermionic higher spin dark matter (HSDM) candidate
which interacts with the Standard Model via a dark U(1) mediator.  For a particular
subclass of interactions, spin-polarized targets may be used for spin determination: The angular
dependence of scatterings can distinguish integer (spin-$s$) vs. half-integer (spin-$s + 1/2$), while
the recoil energy dependence of the signal determines $s$.   
We consider also the signal of a supersymmetric higher
spin dark sector, which suggests a characteristic signal (“SUSY Rilles”) for directional direct detection.
}
\end{center}
$$~~$$
\vfill
Keywords: Dark Matter, higher spin, supersymmetry
\vfill
\clearpage

\tableofcontents

\section{Introduction}
\label{intro}
The existence of dark matter as the predominant form of matter in our universe is established to an incredible degree of certainty \cite{Aghanim:2018eyx}. However, the fundamental properties of the dark matter particle(s), such as the mass and spin, remain unknown. The mass of dark matter could range from $10^{-22}~ {\rm eV}$ to $10^{15}$ grams, while the spin, far from being limited to $s=0, 1/2$, or $1$ as in the Standard Model, could be 3/2 \cite{Ding:2012sm, Chang:2017dvm, Garcia:2020hyo}, 2 \cite{Aoki:2017cnz, Marzola:2017lbt, Armaleo:2019gil, Armaleo:2020yml}, 3 \cite{Asorey:2010zz, Asorey:2015hrz}, or yet higher spin $s$ \cite{Alexander:2020gmv,Criado:2020jkp,Gondolo:2021fqo}.

A prominent search strategy for dark matter is that of direct detection (see e.g., \cite{Lin:2019uvt}). As the name would suggest,  direct detection aims to capture a signature of the dark matter particle as it passes through the experimental apparatus. A variant on this is {\it directional } direct detection \cite{Spergel:1987kx,Mayet:2016zxu}, which additionally captures the angular dependence of scattering events, in part generated by the Earth's motion in the ambient dark matter halo. A further variant is to consider directional direct detection with polarized targets \cite{Chiang:2012ze,Catena:2017wzu}. A small but growing body of work \cite{Catena:2017wzu, Catena:2018uae, Bozorgnia:2011vc} has studied the extent to which direct detection experiments can measure the spin of dark matter, e.g., distinguish a spin-1/2 fermion from a spin-1 boson.

The topic of signatures of particle spin has undergone a renaissance in the context of cosmology, in the form of cosmological collider physics \cite{Arkani-Hamed:2015bza, Chen:2009we, Chen:2015lza,Lee:2016vti} and the closely related cosmological bootstrap \cite{Arkani-Hamed:2018kmz,Baumann:2019oyu,Baumann:2020dch}. The target observables in this context are correlation functions of the primordial curvature perturbation as inferred from the cosmic microwave background anisotropies. 
While most work also in this context has focused on spin-$0$, spin-$1/2$, and spin-$1$, exciting results have been found for spin-$s >1$ , see \cite{Arkani-Hamed:2015bza,Lee:2016vti,Alexander:2019vtb}. In parallel, higher-spin particles as a dark matter candidate has been proposed in \cite{Alexander:2020gmv,Criado:2020jkp} and studied in detail in e.g.~\cite{Gondolo:2021fqo,Criado:2021itq,Dong:2021yak}. The dark matter in this context can be produced gravitationally \cite{Alexander:2020gmv}, or by either freeze-out or freeze-in processes \cite{Criado:2020jkp}.

 In this work we study the phenomenology of higher-spin dark matter within the framework of directional direct detection, focusing on the angular dependence of nuclear recoils. This is done in analogy to analyses of the `cosmological collider' \cite{Arkani-Hamed:2015bza, Chen:2009we, Chen:2015lza,Lee:2016vti,Alexander:2019vtb}, where angular dependences of correlation functions have been developed as a probe of particle spin. Guided by Ref.~\cite{Catena:2017wzu}, we focus our efforts on spin-polarized targets, for which the physical quantity of interest is,
\beq 
\frac{d\Delta R }{dE_Rd\Omega}= \frac{1}{2}\left(\frac{dR(\vec{s}_N)}{dE_R d\Omega} - \frac{dR(-\vec{s}_N)}{dE_R d\Omega}\right), 
\eeq 
where $dR/(dE_R d\Omega)$ is the double-differential cross section and $\vec{s}_N$ is the polarization vector of the target nuclei. We consider a massive spin-$s$ bosonic and spin-$s+1/2$ fermionic dark matter candidate that is coupled to the standard model through a massive spin-1 mediator. We find that the angular dependence $\frac{d\Delta R }{dE_Rd\Omega}$ is a powerful discriminator of bosonic vs. fermionic higher-spins, though limited in its ability to select out a single value of the spin. However, additional information comes from the recoil energy-dependence: we find that scattering rate of a nuclie with a spin-$s$ or spin-$(s+1/2)$ dark matter particle scales with recoil energy as $E^{2s}$, implying to a steep $s$-dependent fall-off of the rate at low energies. Thus, by combining the angular dependence with the energy spectrum, the (higher) spin of dark matter may be identified. 

We then proceed to extend this scenario to study {\it supersymmetric} higher spins \cite{Curtright:1979uz,Kuzenko:1993jp,Kuzenko:1993jq,Gates:2013ska,Gates:2013rka,Gates:2013tka,Koutrolikos:2015lqa,Buchbinder:2015kca,Gates:2017hmb,
Buchbinder:2019esz,Buchbinder:2020yip,Koutrolikos:2020tel}. 
 Supersymmetric higher spins are motivated by superstring theory, wherein higher spin states corresponding to massive string excitations are organized into supersymmetric multiplets. In \cite{Alexander:2019vtb} the authors found characteristic signals of supersymmetric higher spins, so-called SUSY Rilles \cite{Alexander:2019vtb}, in the form of a tetrad of angular dependences of the 3-point correlation function, with each component of the tetrad encoding one field in a higher-spin supermultiplet. On the dark matter side, an interesting possibility studied in detail \cite{Burgess:2021obw,Burgess:2021juk} is that dark matter, being comprised of multiple fields all very weakly coupled to the standard model, may be supersymmetric. Motivated by this, we construct SUSY Rilles from supersymmetric higher-spin dark matter at directional direct detection experiments.  We again find a tetrad of signals; SUSY Rilles for dark matter directional direct detection. 

The structure of the paper is as follows. In section \ref{Sec:CosmoCollider} we discuss the motivation and origins for this analysis in analogy to the Cosmological Collider Physics formalism \cite{Arkani-Hamed:2015bza, Chen:2009we, Chen:2015lza,Lee:2016vti,Alexander:2019vtb}. In Section \ref{review} we review the higher spin dark matter paradigm and the physics of dark matter directional direct detection. In Section \ref{DDD} we construct interactions of higher spin bosons and fermions with standard model nuclei and analytically calculate the double differential recoil rates. We present numerical results in Section \ref{Results} and the extension to supersymmetric higher spins in Section \ref{sec:SUSY}. We conclude with a discussion and point towards future work in Section \ref{sec:discussion}.

\section{Origins In Cosmological Collider Physics}
\label{Sec:CosmoCollider}
 In recent years there has been significant effort dedicated to understanding scattering of scalar fields with fields of arbitrary spin $s$ of $(s+1/2)$, in the context of inflationary cosmology. In this framework, the primordial curvature perturbations' interactions with heavy fields during inflation leave on the non-Gaussianity of the cosmic microwave background. This approach has been deemed the `cosmological collider' and is best understood in terms of the curvature perturbation three-point function, $\langle \zeta\zeta\zeta\rangle$ (see e.g., \cite{Arkani-Hamed:2015bza, Chen:2009we, Chen:2015lza,Lee:2016vti}). In particular, characterizations of the impacts of particles of arbitrary spin on cosmological observables have been made accessible via the cosmological collider formalism. Previous work has investigated the imprints of spin-s bosons \cite{Lee:2016vti}, spin-s+1/2 fermions, as well as the correlated bosonic and fermionic signatures of higher spin supersymmetry \cite{Alexander:2019vtb}. Our present work is motivated from this perspective, so we will briefly explicate the relevant details below. 

The interaction of a spin-s boson with the curvature perturbation, $\zeta$ is given by
\be
\label{HSbosonints}
\mathcal{L}_{int} \supset  \frac{\lambda_s}{\Lambda^{s-3}}\partial_{i_1}
... \partial_{i_s} \zeta \sigma ^{i_1 ... i_s}+ \frac{g_s}
{\Lambda^{s-2}} \dot{\zeta} \partial_{i_1} ... \partial_{i_s} \zeta  \sigma ^{i_1 ... i_s}~,
\ee
where $\sigma$ is the higher spin boson, $\lambda_s$ and $g_s$ are coupling constants, and $\Lambda$ is a UV cutoff. The 3-point function resulting from higher spin boson exchange in the above interaction is \cite{Lee:2016vti},
\begin{align}
\lim_{k_1\ll k_3, \eta\rightarrow0} \frac{\langle\zeta(k_1)\zeta( k_2) \zeta(k_3)\rangle}{\Delta_\zeta^4}&\, = \,  
\alpha_s
 \Delta_\zeta^{-1}\times P_s(\hat{ k}_1\cdot\hat{k}_3)\times {\cal 
I}^{(s)}(\mu_s,c_\pi,k_1,k_3,k_3) \delta(\sum k_i) + ({k}_2\leftrightarrow { k}_3)\, ,
\label{eq:singleSq}
\end{align}
where $ {\cal I}^{(s)}(\mu_s,c_\pi,k_1,k_3,k_3)$ is a complicated function of momenta given in the Appendices of \cite{Lee:2016vti}, $\Delta_\zeta$ is the primordial power spectrum, and $P_{s}$ the Legendre polynomial. One can proceed for a higher spin fermion in an analogous way, with the interaction \cite{Alexander:2019vtb}
\be
\label{LHS1f}
\mathcal{L}  \supset 
\frac{\lambda_s}{\Lambda^{s-1}}\partial_{i_1 ... i_s} 
\zeta \bar{\chi} \psi ^{i_1 ... i_s}
+ \frac{g_s}{\Lambda^{s}}\dot{ \zeta} 
\partial_{i_1 ... i_s} \zeta \bar{\chi} \psi ^{i_1 ... i_s}+c.c.,
\ee
where $\psi$ and $\chi$ are a spin-s+1/2 fermion, and a spin-1/2 fermion, respectively. We similarly obtain an expression for the bispectrum:
\be
\label{zzz2}
\langle \zeta(k_1,0) \zeta(k_2,0) \zeta(k_3,0) \rangle \simeq 
\mathcal{A}_{s+1/2}\frac{\Delta_{\zeta}(k) ^4}{k^6} \mathcal{S}(k_1,k_2,k_3)  
\delta(\sum k_i) \sum_{m=-s} ^s  c_m P_{s} ^m (\hat{k}_1 \cdot \hat{k}_3)  +  
k_2 \leftrightarrow k_3 ,
\ee
where $P_s^m$ are the associated Legendre polynomials, ${\cal S}(k_1,k_2,k_3)$ is a  function 
of the ratios of $k_i$, and $\mathcal{A}$ a complicated prefactor which can be found in detail in \cite{Alexander:2019vtb}. With knowledge of the non-Gaussianity contributions from both the higher spin bosons and higher spin fermions, one can construct the signature from the Y= s+1/2 supermultiplet of supersymmetry, which contains a spectrum of higher spin fields, known as `SUSY rilles':
\bea
\langle \zeta(k_1,0) \zeta(k_2,0) \zeta(k_3,0) \rangle_{\rm HS-SUSY} =  
&&\langle \zeta(k_1,0) \zeta(k_2,0) \zeta(k_3,0) \rangle_{s+1} \nonumber \\
 && + 2 \times \langle \zeta(k_1,0) \zeta(k_2,0) \zeta(k_3,0) \rangle_{s+1/2} 
\nonumber \\
 && + \langle \zeta(k_1,0) \zeta(k_2,0) \zeta(k_3,0) \rangle_{s} \label{zzzHS} \\
  \propto && 
 P_{s+1}(\hat{k}_1 \cdot \hat{k}_3) \; ,\; \displaystyle \sum_{m=-s} ^s 
P_{s} ^m (\hat{k}_1 \cdot \hat{k}_3)\; ,\; P_{s}(\hat{k}_1 \cdot \hat{k}_3)
\eea
In this work, we operate in parallel to the above framework, instead focusing on terrestrial scattering events, e.g., in direct detection experiments. We proceed by calculating a characteristic signatures for higher spin bosons and higher spin fermions, then finally put them together in the context of higher spin supersymmetry. 

\section{Higher Spin Dark Matter}
\label{review}

\subsection{Higher Spin Dark Matter}
\label{HSDM-review}

Despite the role of higher spins in modern theoretical physics (e.g., in nuclear physics, condensed matter physics, string theory and holography), relatively little work ---with exception of \cite{Alexander:2020gmv,Criado:2020jkp,Gondolo:2021fqo,Criado:2021itq,Dong:2021yak, Jain:2021pnk, Jain:2022kwq, Zhang:2021xxa}--- has been done regarding their role in explaining the observed abundance of dark matter. In part this reflects our reliance on the tremendous success of the Standard Model and GR at
explaining a huge amount of phenomena in nature, but mostly it is an indication of the historical challenges on
constructing consistent interactions of higher spin particles. These challenges are often captured by famous no-go results (see \cite{Bekaert:2010hw}). However all of these results rely on a set of assumptions which may prove to be to restrictive. For example, constraining physics at infinity by having a trivial S-matrix ($S=1$) in the presence of massless higher spin particles does not exclude local interactions of higher spins. From this viewpoint the candidacy of higher spins particles as dark matter particles is natural because it immediately explains the \emph{dark} nature of dark matter and the stability of the dark matter particle.
Moreover, none of these no-go results apply to massive higher spins hence they can be used to construct models that produce such dark matter particles.

In \cite{Alexander:2020gmv} it was shown that massive higher spin particles can be produced {\it gravitationally} in sufficient number to explain the observed dark matter abundance.  This scenario gives rise to a large parameter space in both mass and spin that is characterized by the Hubble parameter at the end of cosmic inflation. A characteristic feature of gravitational production of bosonic higher-spins is the dominance of the longitudinal (helicity-0) mode, analogous to the longitudinal component of a massive spin-1 particle. Consistent with this, previous analyses of gravitational production of massive spin-1 (e.g. Refs.~\cite{Ahmed:2020fhc,Kolb:2020fwh}) have found that production of the longitudinal component is parametrically enhanced relative to the transverse components. In the fermionic case, the the production is again dominated by the longitudinal (spin-$1/2$ component), see e.g. \cite{Kolb:2021xfn,Kolb:2021nob} for the spin-$3/2$ case.  Guided by this, we focus primarily on DM scattering involving the longitudinal component of the higher-spins, though the extension to transverse components is straightforward.  We note that contemporaneous work \cite{Falkowski:2020fsu} demonstrated that the freeze-in and freeze-out mechanisms can similarly be ported to a higher-spin dark matter candidate. In all cases, the higher-spin particle is massive, and the theory is understood as a low-energy effective field theory of a broken higher-spin symmetry.

The study of observational signatures of higher-spin dark matter is still in it infancy. In the case of gravitational production of HSDM during inflation \cite{Alexander:2020gmv}, there is a known tell-tale signature of higher-spins in the non-Gaussianity of the Cosmic Microwave Background radiation, in the form of an angular dependence of the three-point correlation function \cite{Lee:2016vti} . While \cite{Alexander:2020gmv} focused on bosons, this can be extended (see Ref.~\cite{Alexander:2019vtb}) to include higher-spin fermions or even to {\it supersymmetric} higher spins, leading to what \cite{Alexander:2019vtb} referred to as ``SUSY Rilles". Motivated by this, in what follows we focus on an analog search strategy for higher spin dark matter, namely directional direct detection searches for dark matter.
\subsection{Directional Direct Detection of Dark Matter}
\label{DDD-review}
A main hope for detecting dark matter particles lies in direct detection experiments. These experiments aim to observe nuclear recoil events from a dark matter particle scattering off a standard model nucleus. Thus far, no direct detection experiment has observed any such event, however there is hope that new and upgraded experiments such as LZ and XENON will be able to confirm their existence using these methods. Directional direct detection is an approach which comes from the observation that the earth moving through the galactic DM halo in a particular direction will induce a preferred direction for nuclear recoil events  \cite{Spergel:1987kx,Mayet:2016zxu}. This suggests that in addition to looking at the differential recoil rate as a function of energy, valuable information can be obtained via the double differential recoil rate, which takes into account the angular dependence of the scattering. Previous studies have shown that it is possible to distinguish the signatures of spin-1/2, spin-0 and spin-1 dark matter using directional direct detection methods \cite{Chiang:2012ze,Catena:2017wzu, Catena:2018uae, Bozorgnia:2011vc}. We will apply the same methods to our study of higher spin dark matter (HSDM). We follow the notation described in \cite{Catena:2018uae}. 

For directional direct detection, the observable quantity of interest is the double differential recoil rate, which is given by \cite{Catena:2018uae}
\begin{align} 
\frac{dR}{dE_Rd\Omega} = \frac{\rho_\chi}{m_\chi m_N}\int d^3v v f(\vec{v}) \frac{d\sigma}{dE_Rd\Omega}.
\label{eq:generaldef}
\end{align} 
Here $\rho_\chi$ is the density of dark matter ($\rho_{\rm{DM}} \simeq0.3$ GeV/cm$^3$) \cite{Baxter:2021pqo}, $m_\chi$ the dark matter mass, $m_N$ the nucleon mass, $v$ is the relative velocity of the incoming DM particle relative to the target nucleus, $f(\vec{v})$ is the velocity distribution of DM in the galactic halo, and $\frac{\rm{d}\sigma}{\rm{d}E_R\rm{d}\Omega}$ is the double differential cross section of the dark matter-nucleus scattering. The double differential cross section can be written as 
\be 
\frac{d\sigma}{d E_Rd\Omega} = \frac{v}{2\pi}\frac{d\sigma}{dE_R}\delta\left(\vec{v}\cdot\hat{q}- \frac{q}{2\mu}\right), 
\ee 
where $d\sigma/dE_R$ is the differential cross section related to the scattering amplitude, $|\mathcal{M}|^2$
via 
\be 
\frac{d\sigma}{dE_R} = \frac{1}{32\pi}\frac{1}{m_\chi^2m_N v^2}|\mathcal{M}|^2,
\label{eq:crosssection}
\ee 
and $\vec q$ is the momentum along the direction of nuclear recoil, related to the recoil energy $E_R$ by $q = \sqrt{2 m_N E_R}$. We assume a Maxwell-Boltzmann velocity distribution function truncated at $v_{esc} = 544$ km/s, with the most probable speed $v_0 = 220$ km/s. For simplicity, we take  $\Vec{v}_e$, the Earth velocity in the galactic rest frame to be $v_e = 232$ km/s \footnote{See \cite{Baxter:2021pqo} for slightly updated numbers.}, defined such that $v_e$ is in the z-direction. We define a set of angular variables via
\bea 
 \hat{q} = (\mathrm{sin}\alpha \mathrm{cos}\beta, \mathrm{sin}\alpha \mathrm{sin}\beta, \mathrm{cos}\alpha), \\
\hat{v} = (\mathrm{sin}\theta \mathrm{cos}\phi, \mathrm{sin}\theta \mathrm{sin}\phi, \mathrm{cos}\theta),
\eea 
where $\hat{v}$ is the direction of the incoming dark matter particle and $\hat{q}$ is the recoil direction of the standard model nucleus. We also define the outgoing DM direction as 
\beq 
\hat{v}^\prime = \frac{v}{v^\prime}\hat{v} + \frac{q}{m_\chi v^\prime}\hat{q}.
\eeq 
The geometry of this system is illustrated in Fig.~\ref{fig:drawings}.

\begin{figure}[ht]
\centering
\begin{minipage}{.49\linewidth}
\centering
       \def\svgwidth{1.0\textwidth}
\begingroup%
  \makeatletter%
  \providecommand\color[2][]{%
    \errmessage{(Inkscape) Color is used for the text in Inkscape, but the package 'color.sty' is not loaded}%
    \renewcommand\color[2][]{}%
  }%
  \providecommand\transparent[1]{%
    \errmessage{(Inkscape) Transparency is used (non-zero) for the text in Inkscape, but the package 'transparent.sty' is not loaded}%
    \renewcommand\transparent[1]{}%
  }%
  \providecommand\rotatebox[2]{#2}%
  \newcommand*\fsize{\dimexpr\f@size pt\relax}%
  \newcommand*\lineheight[1]{\fontsize{\fsize}{#1\fsize}\selectfont}%
  \ifx\svgwidth\undefined%
    \setlength{\unitlength}{134.79306559bp}%
    \ifx\svgscale\undefined%
      \relax%
    \else%
      \setlength{\unitlength}{\unitlength * \real{\svgscale}}%
    \fi%
  \else%
    \setlength{\unitlength}{\svgwidth}%
  \fi%
  \global\let\svgwidth\undefined%
  \global\let\svgscale\undefined%
  \makeatother%
  \begin{picture}(1,0.72232441)%
    \lineheight{1}%
    \setlength\tabcolsep{0pt}%
    \put(0,0){\includegraphics[width=\unitlength,page=1]{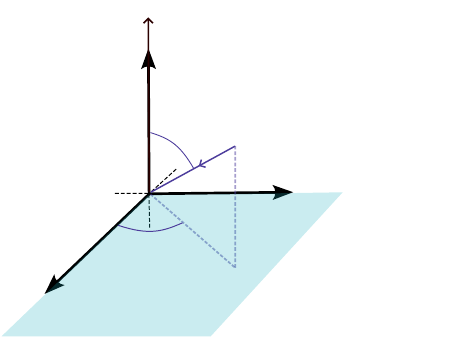}}%
    \put(0.11701215,0.0466601){\makebox(0,0)[lt]{\lineheight{1.25}\smash{\begin{tabular}[t]{l}$\hat{x}$\end{tabular}}}}%
    \put(0.58269275,0.26189946){\makebox(0,0)[lt]{\lineheight{1.25}\smash{\begin{tabular}[t]{l}$\hat{y}$\end{tabular}}}}%
    \put(0.51301925,0.38906417){\makebox(0,0)[lt]{\lineheight{1.25}\smash{\begin{tabular}[t]{l}$\vec{v}$\end{tabular}}}}%
    \put(0.25698841,0.56357727){\makebox(0,0)[lt]{\lineheight{1.25}\smash{\begin{tabular}[t]{l}$\hat{z}$\end{tabular}}}}%
    \put(0.34191998,0.66359595){\makebox(0,0)[lt]{\lineheight{1.25}\smash{\begin{tabular}[t]{l}$\vec{v}_e$\end{tabular}}}}%
    \put(0.27477835,0.18542645){\makebox(0,0)[lt]{\lineheight{1.25}\smash{\begin{tabular}[t]{l}$\phi$\end{tabular}}}}%
    \put(0.36556008,0.42996723){\makebox(0,0)[lt]{\lineheight{1.25}\smash{\begin{tabular}[t]{l}$\theta$\end{tabular}}}}%
  \end{picture}%
\endgroup%

\end{minipage}
\begin{minipage}{.49\linewidth}
\centering
       \def\svgwidth{1.0\textwidth}
       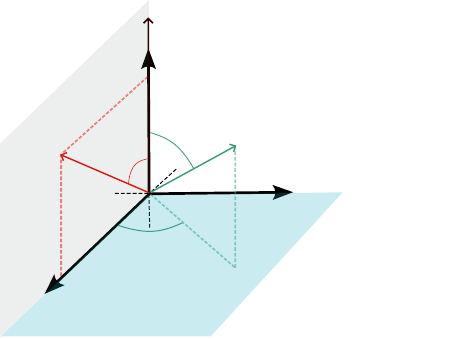
\end{minipage}
\caption{Geometry of the dark matter-nucleon scattering. The incoming dark matter (left panel) has momentum $m_\chi{\vec{v}}$ with direction specified by the angles $(\theta,\phi)$, and the outgoing nucleon has momentum ${\vec{q}}$ with direction specified by the angles $(\alpha,\beta)$. The polarization angle with respect to the velocity of the Earth in the galactic rest frame is taken to be $\vartheta$.}
   \label{fig:drawings}
\end{figure}

Using these definitions, we can write the recoil rate as \cite{Catena:2018uae}
\begin{align}
\frac{d^2R}{dE_R d\Omega} &= \frac{\rho_{\rm{DM}}}{2N\pi m m_N}\int_{-1}^{1}
\mathrm{dcos}\theta \int_0^{2\pi}\mathrm{d}\phi \frac{ \bar{v}^4}{|\hat{v}\cdot \hat{q}|} e^{-(\bar{v} + v_e)^2/v_0^2} \frac{d\sigma}{dE_R}(\bar{v})\Big[\Theta(\bar{v} - v_{min})\nonumber \\ &\Theta((v_{esc} - v_e) - \bar{v}) \Theta((v_{esc}^2 - \bar{v}^2 - v_e^2)/(2\bar{v}v_e)-\cos\theta)\Theta(\bar{v} - (v_{esc}- v_e))\Theta((v_{esc} + v_e) - \bar{v})\Big],
\label{eq:R1}
\end{align}
where N is a constant given by 
\beq 
N = \pi v_0^2\left[\sqrt{\pi}v_0\,\mathrm{erf}\left(\frac{v_{esc}}{v_0}\right) - 2 v_{esc}e^{-v_{esc}^2/v_0^2}\right].
\eeq 
and 
\beq 
\bar{v} = \frac{q}{2\mu(\hat{v}\cdot\hat{q})}.
\eeq 
To obtain this expression, we have split the integral in Eq.~\eqref{eq:generaldef} into two distinct parts, one from $v_{min} - (v_{esc} - v_e)$ and one from $(v_{esc} - v_e) - (v_{esc} + v_e) $, where $v_{min}$ is the minimum velocity for the dark matter to lead to a recoil event. We then performed the integral over $v$ using 
\be 
\delta\left(\vec{v}\cdot \hat{q} - \frac{q}{2\mu}\right) = \frac{\delta(v-\bar{v})}{|\hat{v}\cdot\hat{q}|}. 
\ee 

In order to maximize the sensitivity of the angular recoils to the spin of the dark matter particle, we focus on the particular experimental setup that utilizes a polarized target (see e.g. Refs.~\cite{Chiang:2012ze,Catena:2018uae}). Thus, rather than just calculating the recoil rate, Eq.~\eqref{eq:R1}, we will consider the spin-difference recoil rate, given by, 
\beq 
\frac{d\Delta R }{dE_Rd\Omega}= \frac{1}{2}\left(\frac{dR(\vec{s})}{dE_R d\Omega} - \frac{dR(-\vec{s})}{dE_R d\Omega}\right), 
\eeq 
where $\vec{s}$ is the polarization vector of the target nuclei defined by 
\beq 
\vec{s} = 2\vec{S}_N, 
\eeq 
where $S_N = \Vec{\sigma}_N/2$, with $\sigma_N$ the Pauli matrices, and we take the polarization direction to be in the $x-z$ plane such that
\begin{equation}
  \hat{s} = (\sin\vartheta, 0, \cos\vartheta).
\end{equation}

\section{
Higher-Spin - Standard Model Nucleon Scattering}
\label{DDD}

We will consider the scenario in which a HSDM particle interacts with a standard model nucleus via a spin-1 mediator. To this end, we consider a complex higher spin field, which we compose out of two real spin-$s$ fields, e.g.
\be 
\sigma_{\mu_1...\mu_s} = \sigma^1_{\mu_1...\mu_s} + i\sigma^2_{\mu_1...\mu_s}.
\ee 
There are several classes of interactions one can write. In what follows we focus on a particular type of interaction in
which all Lorentz indices of the higher-spin fields are contracted with derivatives acting on the mediator. We emphasize that the interactions we present below are particular examples out of the set of all possible interactions, chosen such that we can compare with existing results
regarding the directional direct detection of spin-1 and spin-1/2 dark matter found in \cite{Catena:2018uae}. We work within an effective field theory framework such that the interactions we present are valid up to a UV cutoff scale, $\Lambda$. We further focus on the scenario in which the dark matter is gravitationally produced, as discussed in \cite{Alexander:2020gmv}. Note that the discussion in \cite{Alexander:2020gmv} focused soley on bosons, due to the lack of knowledge of fermionic higher spins in de Sitter space and inflationary spacetimes. However, complete knowledge of the primordial production in de Sitter space is not required to compute simple Feynman diagrams. We will proceed here with the fermionic case in analogy to \cite{Alexander:2019vtb}, motivated by higher spin supersymmetry and `SUSY Rilles.'

\subsection{Higher Spin Bosons}\label{hsb}
First, let us consider a bosonic spin-$s$ HSDM particle which interact with a standard model nucleus via a vector mediator. We work in analogy with the spin-1 analysis presented in \cite{Catena:2018uae}, and consider a subset of possible interactions within the framework of a low energy effective field theory. These interactions were identified in \cite{Catena:2018uae} and lead to a  characteristic angular dependence. We focus on the interaction:
 \begin{align}
 \label{eq:LintBoson}
\mathcal{L}^{(s)}_{int} &= -\frac{1}{\Lambda^{2s}}\left[(\partial_{\nu}\sigma^{\mu_1...\mu_s} )\sigma^{\dagger \mu_{s+1}...\mu_{2s}}- (\partial_{\nu}\sigma^{\dagger\mu_{s+1}...\mu_{2s}}(p'))\sigma^{\mu_1...\mu_s}\right]\partial_{\mu_1...\mu_{2s}}G^\nu \nonumber \\
& - h_3 G_\mu \bar{N}\gamma^\mu N - h_4G_\mu \bar{N}\gamma^\mu\gamma_5 N, 
\end{align}
where $\Lambda, h_3$, and $h_4$ are coupling constants, $\sigma$ is a complex spin-$s$ field, $N$ is a standard model nucleus  and $G$ is the vector mediator.  

A higher spin bosonic field, $\sigma$, can be decomposed in constant time slices in the following way \cite{Lee:2016vti}: 
\be 
\sigma_{i_1...i_n0...0} = \sum_\lambda \sigma^\lambda_{n,s}\varepsilon^\lambda_{i_1...i_n}, 
\ee 
where $\sigma^\lambda_{n,s}$ are the HS mode functions, $\varepsilon$ are the spin-$s$ polarization tensors. Here, $s$ refers to the spin, $n$ is the `spatial spin' and $\lambda$ is the helicity of the field. Note that there are $(n-s)$ number of temporal $(0)$ indices. Previous work has shown that the $\sigma^0_{s,s}$ component is the dominant helicity state from gravitationally produced HSDM \cite{Alexander:2020gmv}. In what follows we focus solely on the $\lambda = 0$ component of $\sigma$, and note the numerical results are unchanged when the other helicity states are included.  

A key property of the spin-$s$ polarization tensors is that when they are contracted with factors of momenta, one obtains a characteristic Legendre polynomial dependence:
\be 
p^{\mu_1...\mu_s}\varepsilon_{\mu_1...\mu_s}(k) = |p|^{s}P_s(\hat{p}\cdot\hat{k}), 
\ee 
where $P_s$ is a Legendre polynomial. With this characteristic angular dependence, one is able to distinguish a genuine higher spin field from a scalar field with $s$ derivatives. The latter would simply lead to a factor of $|p|^s$. For further details regarding this decomposition and the spin-$s$ polarization tensors, see Appendix \ref{spinsPol} or \cite{Lee:2016vti}. With this knowledge, we can write down the matrix element for this interaction as: 
\be 
i\mathcal{M} = -i\frac{ |q|^{2s}}{\Lambda^{2s}m_G^2}P_s(\hat{p}\cdot \hat{q})P_s(\hat{p}'\cdot \hat{q})\left(p^\mu  + p'^\mu \right) \bar{u}_N\gamma_\mu (h_3 + \gamma_5h_4)u_N),
\ee 
where $p$ and $p'$ are the ingoing and outgoing HSDM momenta, respectively, defined with $\vec{p} = m_\chi\vec{v}$ and $\vec{p}' = m_\chi\vec{v}'$. From this expression, we take the non-relativistic limit to obtain: 
\be 
|\bar{\mathcal{M}}|^2 = \frac{16m_N^2 m_\chi^2}{m_G^4}\frac{|q|^{4s}}{\Lambda^{4s}} P_s(\hat{v}\cdot \hat{q})^2 P_s(\hat{v}'\cdot\hat{q})^2 \left[h_3^2 - h_3h_4 \left(1 - \frac{m_\chi}{m_N}\right)\vec{v}\cdot\vec{s} - h_3h_4\left(1 + \frac{m_\chi}{m_N}\right)\vec{v}'\cdot\vec{s}\right]. 
\ee
For technical details of this computation, see Appendix \ref{Mcalc} . Then, using Eqs.~\eqref{eq:crosssection} and~\eqref{eq:R1}, we find the 
spin-differenced double differential recoil rate to be 
\begin{align}
\frac{d\Delta R}{dE_rd\Omega} &= \mathcal{F}_b\int_{-1}^{1}d\cos\theta \int_0^{2\pi}d\phi
P_s(\hat{v}\cdot\hat{q})^2P_s(\hat{v}'\cdot\hat{q})^2 
\frac{\bar{v}^2}{|\hat{v}\cdot\hat{q}|}e^{-(\bar{v}^2 + v_e^2 + 2\bar{v}v_e\cos\theta)/v_0^2}\nonumber\\
&\times \left[-h_3h_4\left(1 - \frac{m_\chi}{m_N}\right)\Vec{v}\cdot\Vec{s} - h_3h_4\left(1 + \frac{m_\chi}{m_N}\right)\Vec{v}'\cdot\Vec{s}\right]\Big\{\Theta(\bar{v} - v_{min})\Theta\Big[(v_{esc} - v_{e}) - \bar{v}\Big] \nonumber\\
&\times\Theta\Big[(v_{esc}^2 - \bar{v}^2 - v_e^2)/(2\bar{v}v_e) - \cos\theta\Big]
\Theta\Big[\bar{v} - (v_{esc} - v_e)\Big]\Theta\Big[(v_{esc} + v_e) - \bar{v}\Big]\Big\}, 
\label{eq:recoilB}
\end{align}
where $\mathcal{F}_b$ is an overall prefactor given by 
\be 
\mathcal{F}_b = \frac{\rho_\chi}{64\pi^2 N m_\chi m_G^4}\left(\frac{q}{\Lambda}\right)^{4s} . 
\ee 
We note that $q$ is related to the recoil energy by $q = \sqrt{2 m_N E_R}$, thus the above indicates a strong dependence of the rate on recoil energy.
The expressions above are analogous to that found in \cite{Catena:2018uae} for spin-$1$ dark matter, with an additional spin dependence arises via the Legendre polynomials, $P_s$, as well as in the overall prefactor and momentum dependence.

The advantage of interaction \eqref{eq:LintBoson} is that it provides
the clearest spin dependent effects, in both the angular dependence and momentum dependence.
This becomes apparent if one considers other types of interaction. For example
a higher spin mass-like term such as:
\be 
\mathcal{L}_{int} \propto  \sigma_{\mu_1...\mu_s}\sigma^{\dagger \mu_1...\mu_s}\partial_\nu G^\nu. 
\ee 
due to the self-contraction identities of the spin-$s$ polarization tensors (see Appendix \ref{spinsPol}), leads to an overall $s$-dependent pre-factor in the recoil rate, but not the $P_s$ angular dependence or $s$-dependent momentum falloff. More generally, the set of all cubic interaction terms will have some of the indices contracted among the fields and the remaining indices will be contracted with derivatives that can be distributed in many ways among the three fields. Hence, the recoil rate would have a combination of momentum and angular dependences
governed by a distribution of effective spin values which are all less than the true spin value of the dark matter particle.

\subsection{Higher Spin Fermions}\label{hsf}

We now turn our attention to fermionic higher spin dark matter, following the same procedure as in the previous section. We consider an interaction of the form:
\be 
\label{eq:LintFermion}
\mathcal{L}^{(s+1/2)}_{int}= -\lambda_3\bar{N}\gamma^\mu N G_\mu - \lambda_4\bar{N}\gamma^\mu\gamma_5N G_\mu - \bar{\psi}^{\mu_1...\mu_s}\gamma^\nu\left(\frac{1}{\Lambda_3^{2s}} + \gamma_5\frac{1}{\Lambda_4^{2s}}\right)\psi^{\mu_{s+1}...\mu_{2s}}\partial_{\mu_1...\mu_{2s}}G_\nu, 
\ee
where once again we consider a HS particle, in this case a fermion, interacting with a standard model nucleus via a spin-1 mediator. Here $\lambda_3, \lambda_4, \Lambda_3$, and $\Lambda_4$ are all coupling constants, $\psi$ is a spin-$(s+1/2)$ fermion, and $N$ and $G_\mu$ are once again the standard model nucleus and mediator, respectively. Now, to proceed as before, we must consider a decomposition of the HS fermion. In analogy to the decomposition of a bosonic higher spin in terms of the polarization tensor, we can can decompose the HS fermion in the following way 
\be 
\psi_{\mu_1...\mu_n0...0} = \displaystyle \sum _ {\lambda=-s} ^{+s} c_{\lambda} u^\lambda(p)\epsilon^\lambda_{\mu_1...\mu_n}, 
\label{fermiondecomp}
\ee 
where $\epsilon^\lambda_{\mu_1...\mu_n}$ is the usual spin-$s$ polarization tensor, $u(p)^\lambda$ is a spin-$1/2$ eigenspinor, and $c_\lambda$ is a  Clebsch-Gordan coefficient (see e.g.\cite{Moroi:1995fs}).  Once again, we consider only the $\lambda=0$ dominant helicity mode from gravitational production as the dark matter and rescale to absorb $c_\lambda$; as with the bosons, the numerical results are insensitive to including the additional helicity states. With this decomposition, the matrix element is 
\be 
-i\mathcal{M} = -\frac{i}{m_G^2} P_s(\hat{p}\cdot\hat{q})P_s(\hat{p}'\cdot\hat{q}) |q|^{2s} \bar{u}(p)\gamma^\nu \left(\frac{1}{\Lambda_3^{2s}} + \gamma_5\frac{1}{\Lambda_4^{2s}}\right)u(p'). 
\ee 
Proceeding as before and taking the non-relativistic limit, we find
\begin{align}
    |\bar{\mathcal{M}}|^2 &= \frac{16m_\chi^2m_N^2}{m_G^2}|q|^{4s}P_s(\hat{p}\cdot\hat{q})^2 P_s(\hat{p}'\cdot\hat{q})^2\Bigg\{\Lambda_3^{4s} h_3^2 + 3\Lambda_4^{4s}h_4^2 \nonumber \\
    &- \left[\Lambda_3^{4s}h_3h_4\left(1- \frac{m_\chi}{m_N}\right) + \Lambda_4^{4s}h_3h_4\left(1+ \frac{m_\chi}{m_N}\right) + 2\Lambda_3^{2s}\Lambda_4^{2s}h_4^2\right]\vec{v}\cdot\vec{s}\nonumber \\
    &- \left[\Lambda_3^{4s}h_3h_4\left(1+ \frac{m_\chi}{m_N}\right) + \Lambda_4^{4s}h_3h_4\left(1- \frac{m_\chi}{m_N}\right) - 2\Lambda_3^{2s}\Lambda_4^{2s}h_4^2\right]\vec{v}'\cdot\vec{s}\Bigg\}, 
\end{align}
which allows us to write the double differential recoil rate as 
\begin{align}
\frac{d\Delta R}{dE_rd\Omega} &= \mathcal{F}_f\int_{-1}^{1}d\cos\theta \int_0^{2\pi}d\phi
P_s(\hat{q}\cdot\hat{v})^2P_s(\hat{q}\cdot\hat{v}')^2 
\frac{\bar{v}^2}{|\hat{v}\cdot\hat{q}|}e^{-(\bar{v}^2 + v_e^2 + 2\bar{v}v_e\cos\theta)/v_0^2}\nonumber\\
&\times-\Bigg\{ \left[\Lambda_3^{4s}h_3h_4\left(1- \frac{m_\chi}{m_N}\right) + \Lambda_4^{4s}h_3h_4\left(1+ \frac{m_\chi}{m_N}\right) + 2\Lambda_3^{2s}\Lambda_4^{2s}h_4^2\right]\vec{v}\cdot\vec{s}\nonumber \\
    &- \left[\Lambda_3^{4s}h_3h_4\left(1+ \frac{m_\chi}{m_N}\right) + \Lambda_4^{4s}h_3h_4\left(1- \frac{m_\chi}{m_N}\right) - 2\Lambda_3^{2s}\Lambda_4^{2s}h_4^2\right]\vec{v}'\cdot\vec{s}\Bigg\}\Big\{\Theta(\bar{v} - v_{min})\Theta[(v_{esc} - v_{e}) - \bar{v}] \nonumber\\
&\times\Theta[(v_{esc}^2 - \bar{v}^2 - v_e^2)/(2\bar{v}v_e) - \cos\theta]
\Theta[\bar{v} - (v_{esc} - v_e)]\Theta[(v_{esc} + v_e) - \bar{v}]\Big\}, 
\label{eq:recoilF}
\end{align}
where , $\mathcal{F}_f$ is a numerical prefactor defined as 
\be 
\mathcal{F}_f = \frac{\rho_\chi}{64\pi^2 N m_\chi m_G^4}q^{4s}. 
\ee 

\subsection{Imprint of Higher-Spin}
At this point we may already identify the imprint of higher-spin dark matter on the scattering with nuclei, in particular on the double differential recoil rate. We note the following features:
\begin{enumerate}
    \item {\bf Angular Dependence:} The angular dependence distinguishes integer and half-integer spin. The higher spin case enhances this distinction in comparison with the spin-$1$ and spin-$1/2$ results of \cite{Catena:2018uae} due to the measure of angular integration, which now involves Legendre polynomials. The measure is $d\Omega 
P_s(\hat{q}\cdot\hat{v})^2P_s(\hat{q}\cdot\hat{v}')^2$, where $\hat{q}\cdot\hat{v}$ is the angle between the outgoing nucleus and ingoing DM, respectively, and $\hat{q}\cdot\hat{v}'$ is the angle between the outgoing nucleus and outgoing DM. Again, recall that the Legendre polynomial dependence entering in the measure is a property of interactions of an intrinsic higher spin field and cannot arise from a scalar or a vector with derivatives. 
    \item {\bf Recoil Energy Dependence:} The bosonic and fermionic cases both exhibit an $s$-dependent fall-off of the rate at low recoil energy as $\frac{d\Delta R}{dE_rd\Omega} \propto q^{4s}$, where $q$ is the momentum of the outgoing nuclei, related to the recoil energy $E_R$ by $q = \sqrt{2 m_N E_R}$, 
\end{enumerate} 
Based on these features we conclude that the angular and energy dependence of the recoil rate yield complementary information which can be used to determine the spin of massive dark matter particles.
Moreover, we note that the inclusion of the transverse components of higher-spin fields, i.e., components other than the $\lambda=0$ longitudinal mode, will introduce further spin-dependent modifications to the measure of the angular integration. However these modifications will be identical for bosons and fermions and thus would not add to the overall ability to discriminate spin.
\section{(Higher) Spin Determination with Directional Direct Detection}
\label{Results}
We can now numerically investigate the spin-dependent recoil behavior discussed in the previous section. Using Mathematica, we numericallly integrate the expressions in Eqs.~\eqref{eq:recoilB} and~\eqref{eq:recoilF} to obtain the double differential recoil rate as a function of the recoil angles $\alpha$ and $\beta$. For the masses, we take $m_\chi = 1000$ GeV, $m_N= m_G = 100$ GeV. For the couplings, we take $h_3 = -h_4 = 1/2$, and $\Lambda_3 = -\Lambda_4 = \frac{\Lambda}{\sqrt{2}}$, which we leave as a free parameter in the overall scaling of the signal amplitude. We also take $|\vec{s}| = 1$ and $\vartheta = \pi/2$. In what follows, we scale all plots to their maximum amplitude such that the range of differences in the double differential recoil rate goes from -1 to 1, and further normalize to the rate at a reference momentum $q_0$. We define the rate normalized in this way as 
\be 
\left(\frac{\Delta dR}{dE_Rd\Omega}\right)_{norm} = \frac{\Delta dR/dE_Rd\Omega|_q}{\Delta dR/dE_Rd\Omega|_{q_0}}, 
\ee 
where $q_0$ is a reference momentum. In the figures below we choose $q_0$ such that the maximum value of $\left(\frac{\Delta dR}{dE_Rd\Omega}\right)_{norm} $ for $s = 5/2$ is normalized to one. 
We show our results in this way in order to highlight the angular and energy dependence of the nuclear recoils.

First, let us consider the angular dependence of the difference in the double differential recoil rate. In Figure \ref{fig:recoilmain}  we show the spin-polarized difference in the recoil rate in the space of recoil angles $(\alpha,\beta)$. We take the recoil energy to be $E_R = 5$ KeV. We show the result for the both a higher-spin boson ($s=3$) and higher-spin fermion ($s = 5/2$).
From Fig.~\ref{fig:recoilmain} one may appreciate a clear distinction between the bosonic and fermionic cases. For $s = 3$ there is a characteristic  angular feature near $\alpha = \pi/3$. For the fermionic $s=5/2$ case, this same angular dependence exists, but there is now an additional feature on the right hand side that clearly delineates the two cases. This distinction is in agreement with with \cite{Catena:2018uae} for the difference between spin-$1$ and spin-$1/2$ dark matter.

\begin{figure}[h!]
    \centering
    \includegraphics[scale=0.35]{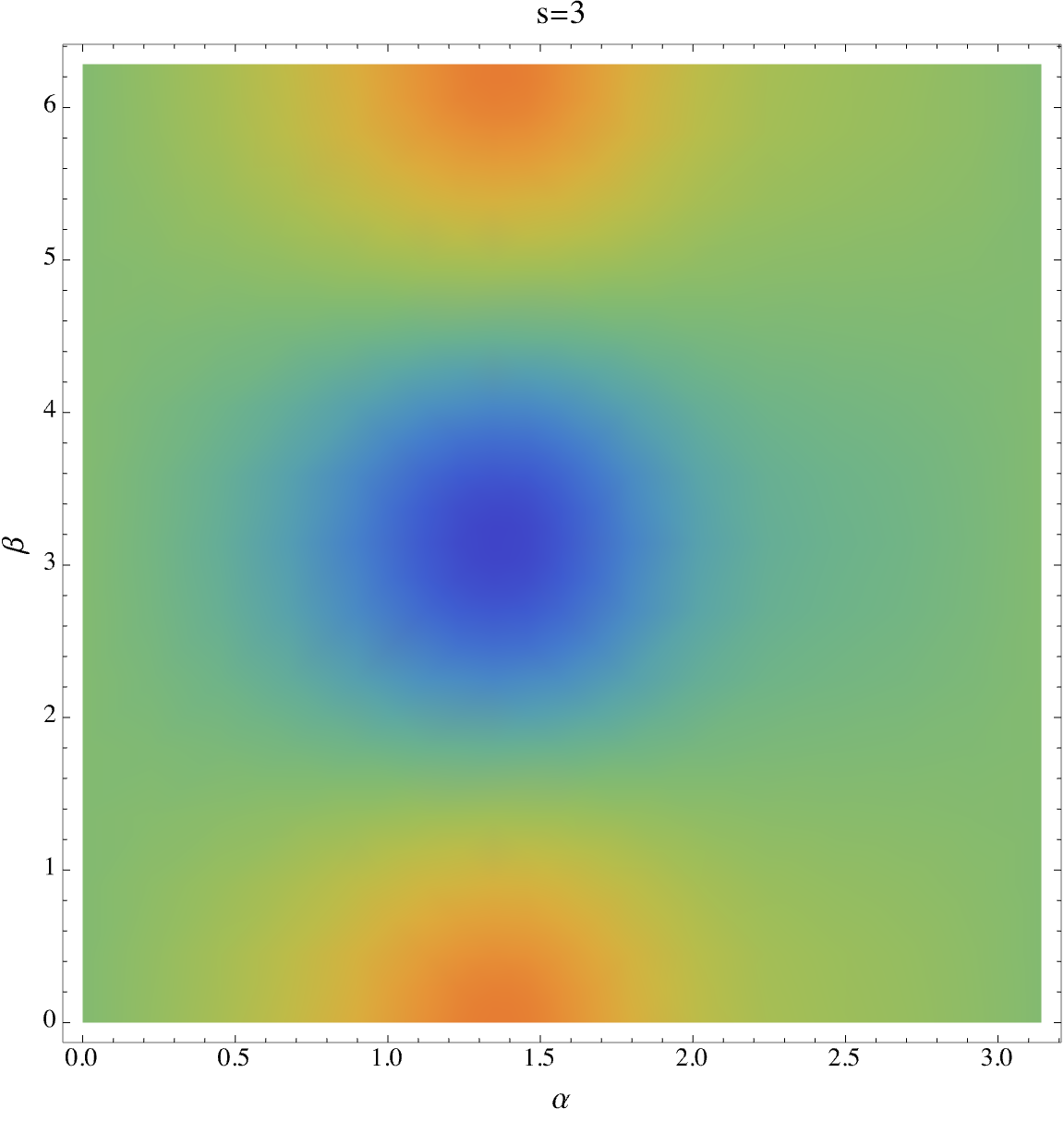}
    \includegraphics[scale=0.35]{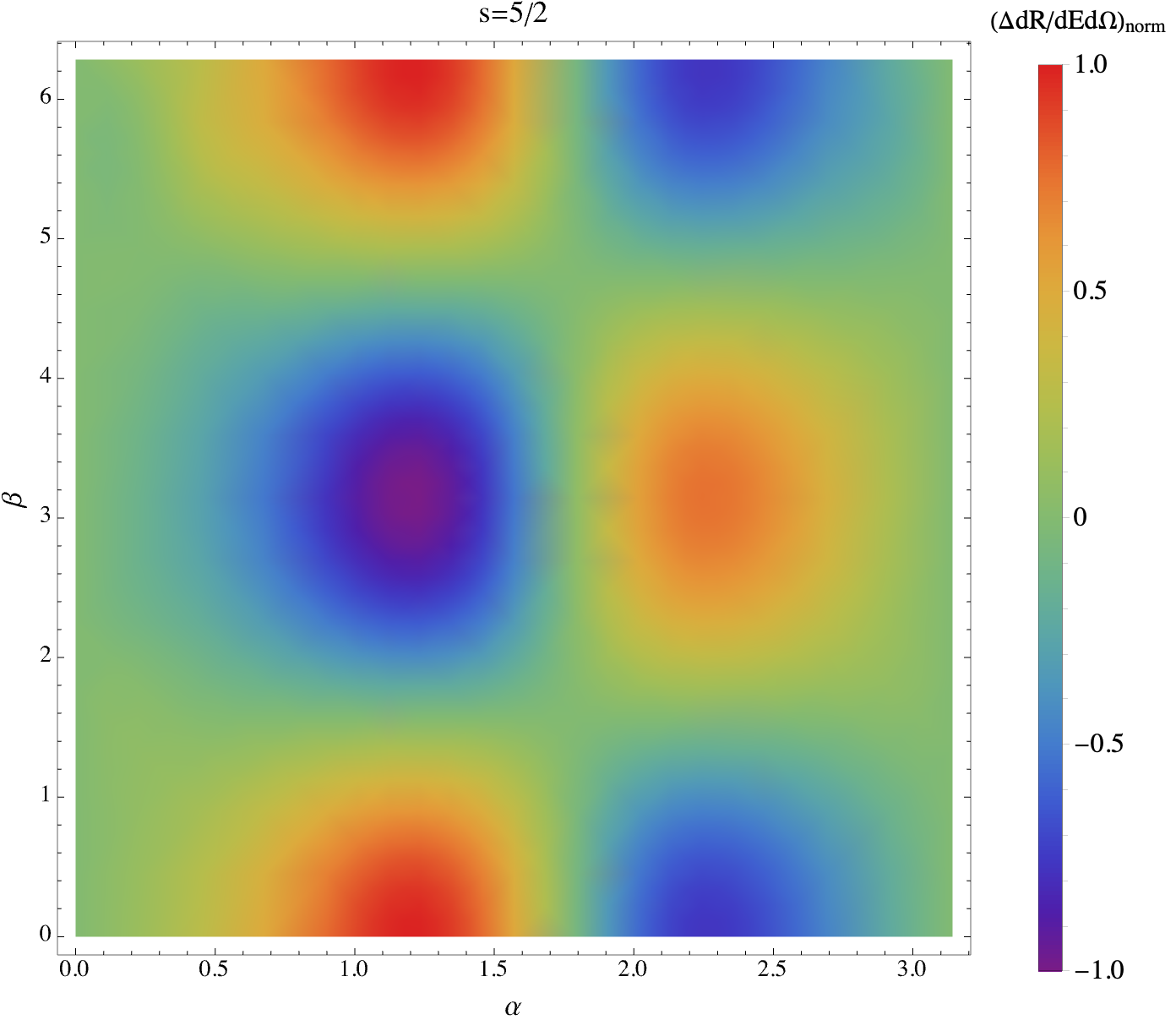}
    \caption{The spin-polarized difference in the differential recoil rate for a spin-3 HSDM boson and a spin-5/2 HSDM fermion shown in the $\alpha,\beta$ plane. The recoil energy is set to $5$ KeV. Note that we enhance the $s=3$ rate by a factor of ten to better show the angular features in comparison to the $s=5/2$ case.}
    \label{fig:recoilmain}
\end{figure}

Remarkably, the angular dependence is unchanged when one generalizes from spin-$3$ and spin-$5/2$ to arbitrary integer spin $s$ and half-integer spin $s+1/2$; i.e., the angular dependence shown in Fig.~\ref{fig:recoilmain} is {\it universal} for higher spin bosons and fermions. This is consistent with the $s=1$ and $s=1/2$ bosonic and fermionic cases as well. We demonstrate this universality in Fig.~\ref{fig:recoilmainSlice}, where we consider a representative case and compare the angular dependence of $s=3$ to $s=1,4,8,12$. We display the root mean squared difference of $s=1,4,8,12$ relative to $s=3$. Away from the zero crossings (which are common across all $s$), we find fractional differences of ${\cal O}(0.01\%)$, which is of order the numerical precision with which the numerical integrations have been performed. This indicates that the angular features are indeed unchanged as one increases the spin. The same result holds for the fermionic case.

\begin{figure}[h!]
    \centering
 \includegraphics[width=0.49\textwidth]{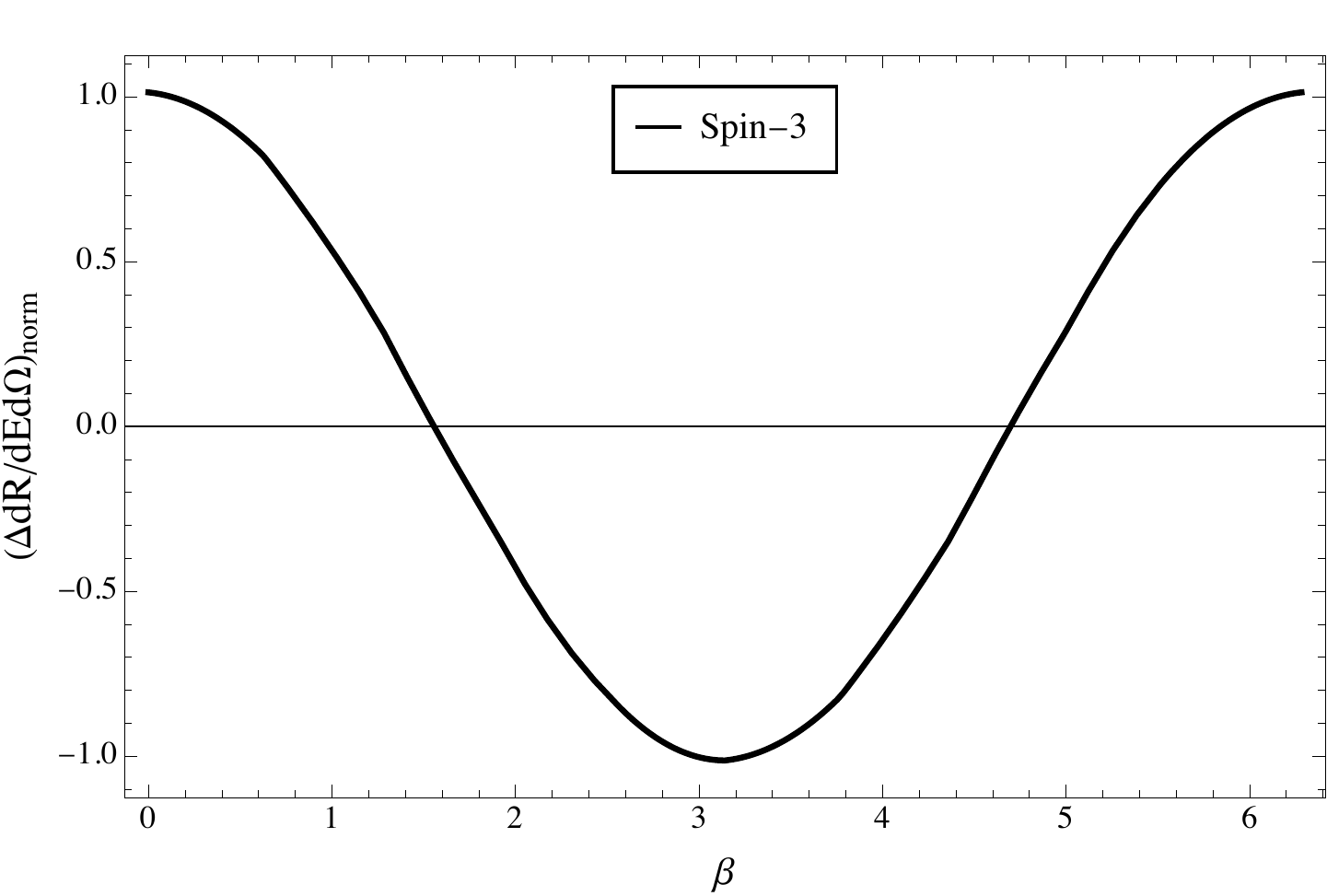}
    \includegraphics[width=.49\textwidth]{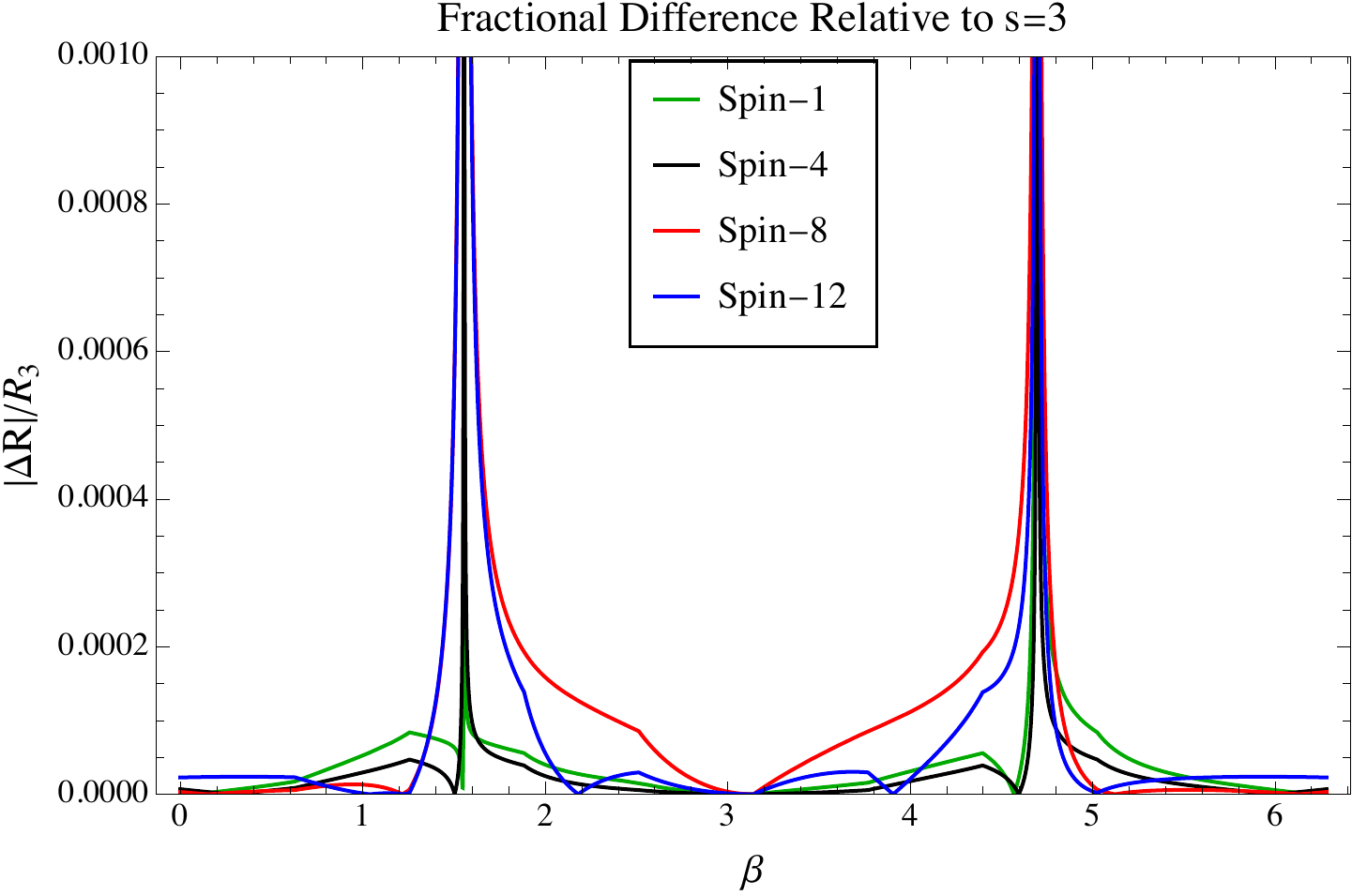}
\caption{The angular ($\beta$) dependence at fixed $\alpha=\pi/3$ for a higher spin boson with $s=3$ (left panel) and the root-mean-squared fractional difference with respect to $s=3$ in the case of $s=1,4,8,12$ (right panel). For each spin we normalize to the value of the rate at $\beta=\pi$, in order to extract the angular dependence. From the small fractional difference (which is ${\cal O}(.01\%)$ away from the zero crossings, on par with the numerical precision of the results) one may  appreciate that the angular-dependence of bosons is {\it universal} across spin. Similar conclusions apply to higher spin fermions.  }
    \label{fig:recoilmainSlice}
\end{figure}

The distinction between fermions and bosons is also manifest when scanning over recoil energy. In Figure \ref{fig:RateEnergy} we plot the recoil rate as a function of energy and recoil angle $\alpha$ for both $s=3$ and $s=5/2$, taking $\beta = \pi$. We can see again that there is a clear deviation between the two, particularly at low energies. This marks another distinction between the spin-$s$ and spin-$(s+1/2)$ dark matter recoils in energy space.

We further note that the characteristic angular features found for higher spin bosons and fermions are the same as those found for the spin-1 and spin-1/2 case in \cite{Catena:2018uae}. While one might expect that increasing the spin should change the angular features due to the presence of the $P_s$ factors in the recoil rate, we note that the double differential recoil rate is an \textit{integrated} effect. Thus, although there are small angular distinctions between different spin values, they are integrated over and washed out, such that the dominant angular contribution arises from the spin-independent part of the rate. From these results one may infer, that regardless of the spin, one can differentiate between bosonic and fermionic dark matter using directional detection with polarized targets. This extends the result found in \cite{Catena:2018uae} from spin-$1$ and spin-$1/2$ to arbitrary spin-$s$ and spin-$(s+1/2)$. 

\begin{figure}[h!]
    \centering
    \includegraphics[scale=0.35]{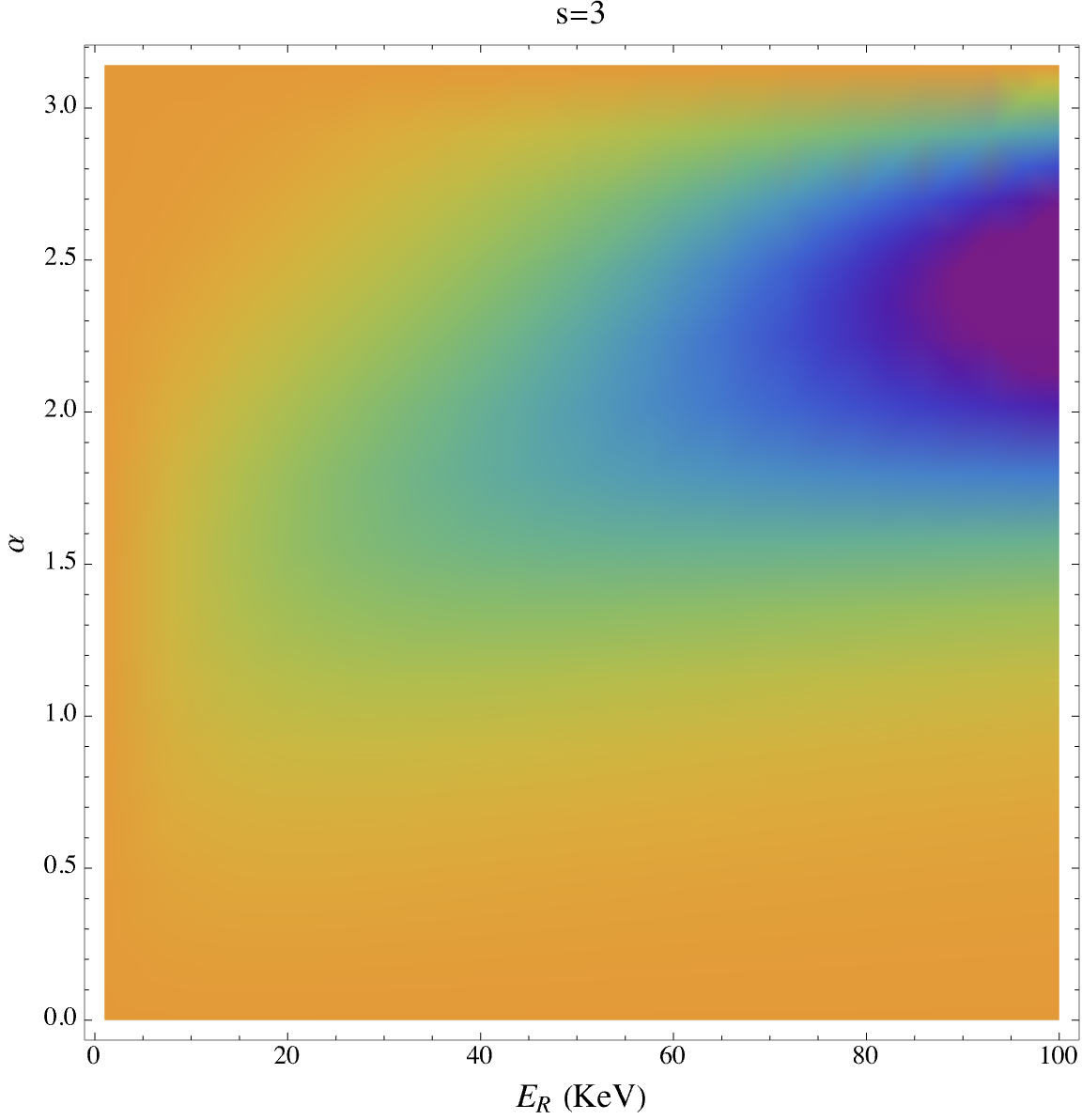}
\includegraphics[scale=0.35]{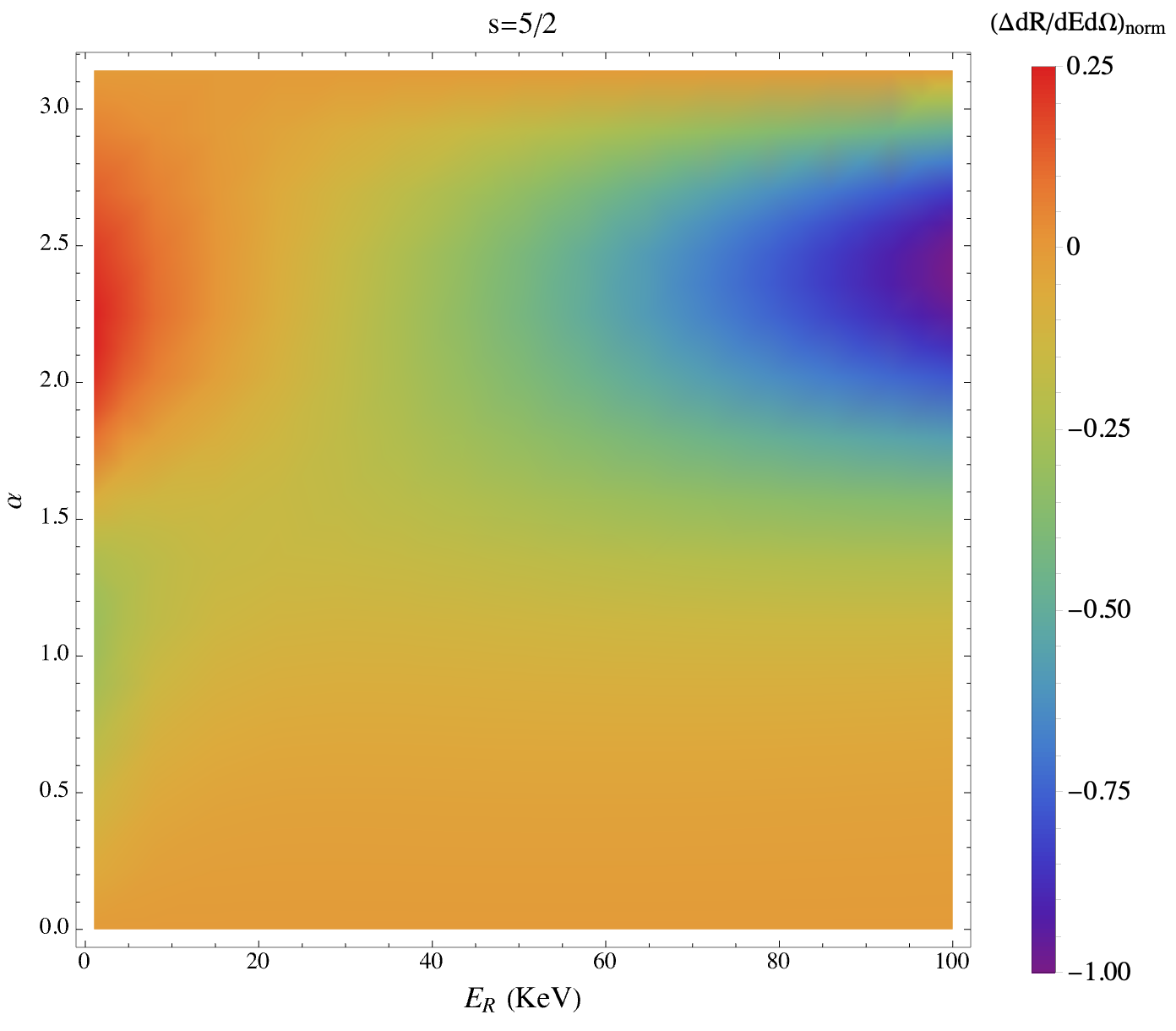}
    \caption{The spin-polarized difference in the differential recoil rate for a spin-3 HSDM boson and a spin-5/2 HSDM fermion as a function of energy and the recoil angle $\alpha$. We take $\beta = \pi$.} 
    \label{fig:RateEnergy}
\end{figure}

To extract the particular spin $s$, and not just the bosonic or fermionic nature of the HSDM, one may utilize the energy dependence.
In Figure \ref{fig:recoilbetapi}  we show a slice of the $\alpha, \beta$ parameter space and set $\beta = \pi$. We show the recoil rate as a function of $\alpha$ in Figure \ref{fig:recoilbetapi} for various values of the outgoing momentum $q$, related to the recoil energy as $q = \sqrt{2 m_N E_R}$, normalized to a reference value $q_0$.  We recall that the rate is proportional to $q^{4s}$ so we expect that the overall rate will be highly sensitive to the recoil energy. 
Here, we can clearly see the distinction in features between the bosonic and fermionic cases. We also see that there is a difference in amplitude of the recoil rate, which encodes the $s$ dependence. Formally, the $s$-dependence and exact scaling encoded in the recoil rate is also dependent on the coupling, $\Lambda$. However, from the above we see that for any given spin, one can determine the value of $s$ by making measurements at at least two different reference momenta. Then, one can compare the amplitudes of the rate of each in order to extract $s$.  %

\begin{figure}[htb]
    \centering
    \includegraphics[scale=.4]{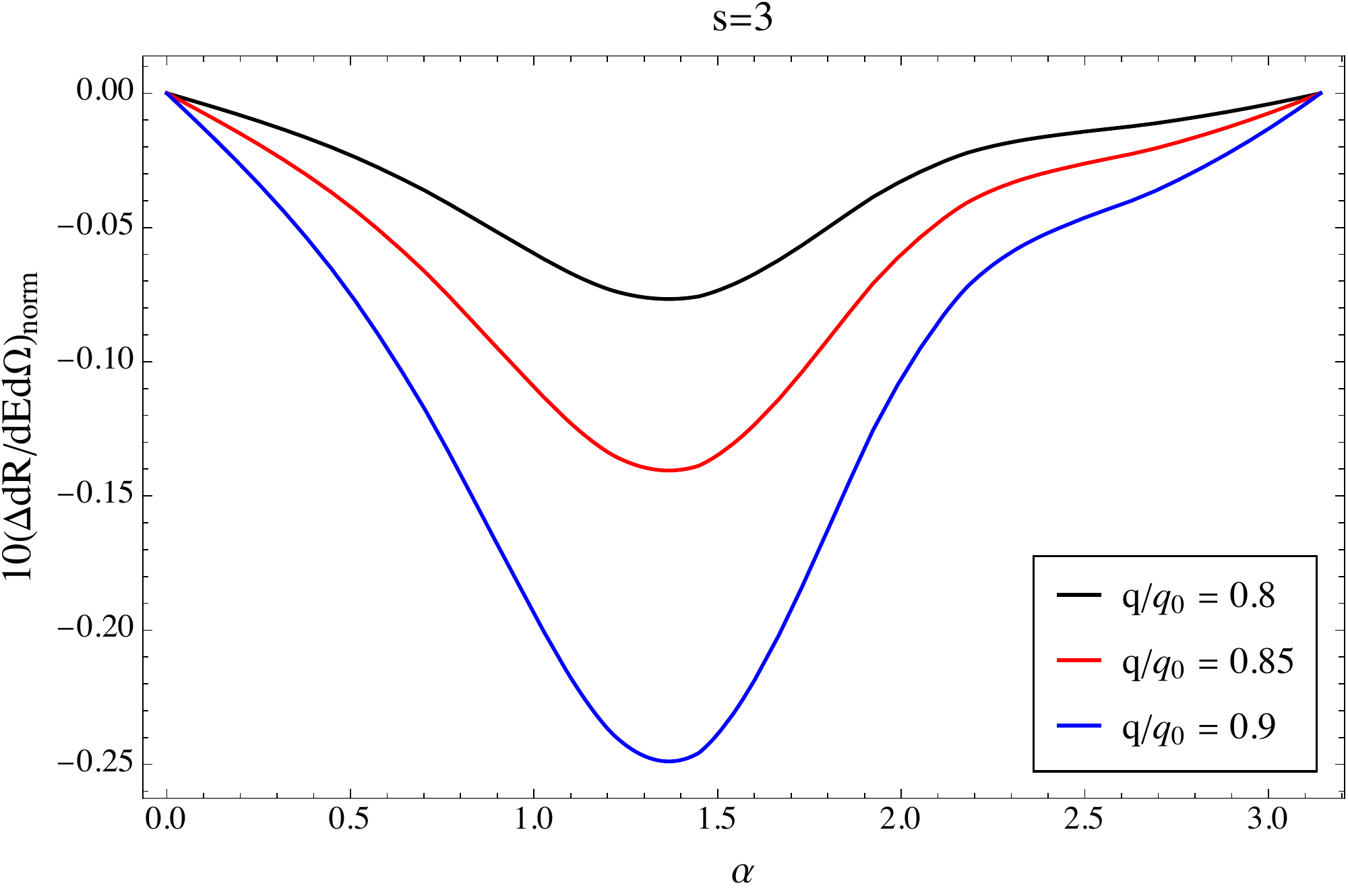}
    \includegraphics[scale=.4]{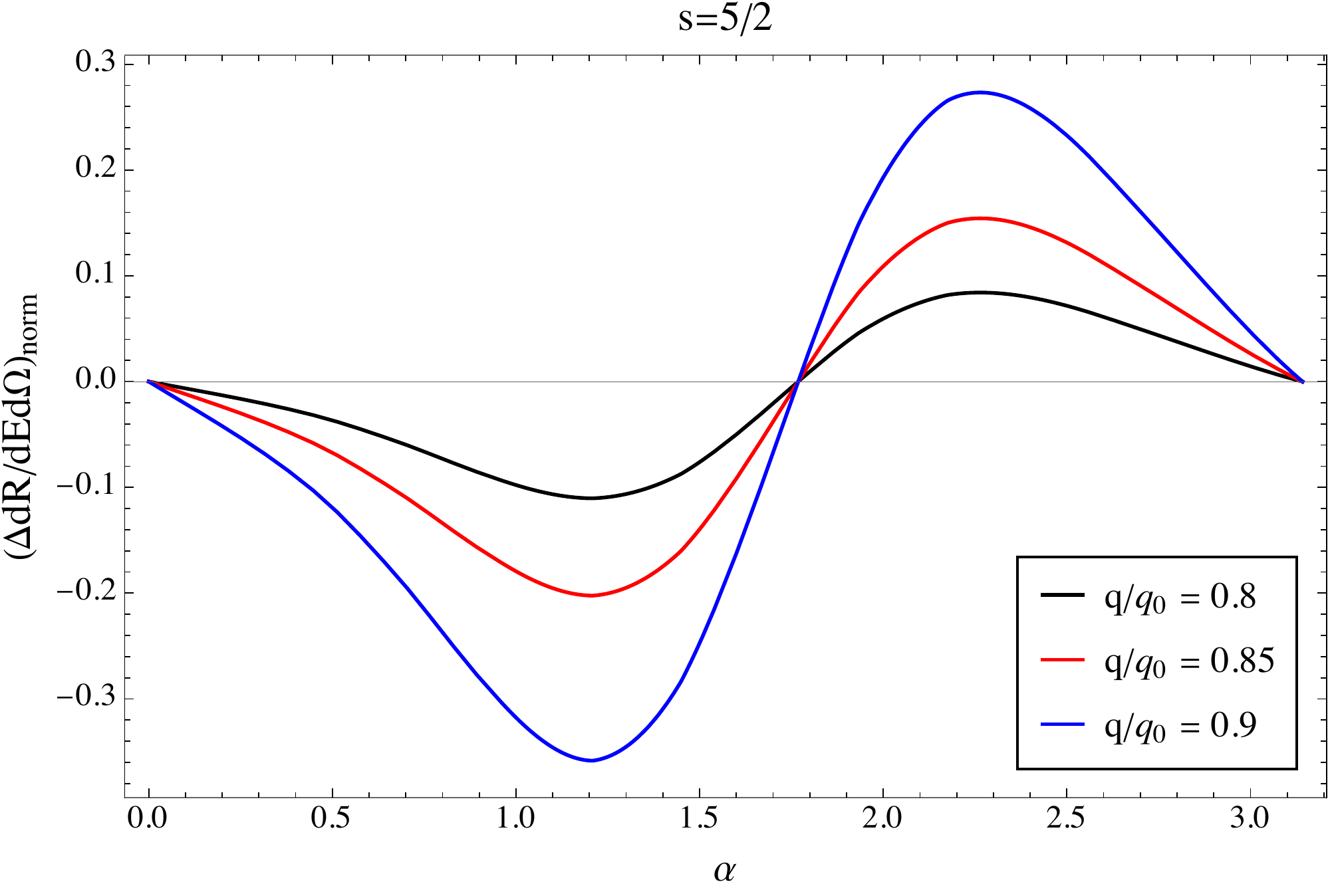}
    \caption{The spin-polarized difference in the differential recoil rate for a spin-3 HSDM boson and a spin-5/2 HSDM fermion at $\beta=\pi$ for various values of the outgoing nuclei momentum $q$, related to the recoil energy as $q = \sqrt{2 m_N E_R}$, normalized to a reference momentum $q_0$. 
    }
    \label{fig:recoilbetapi}
\end{figure}
Lastly, one might wonder how the dark matter mass might influence the above discussion. We find that considering heavier masses in the superheavy range ($10^3 \lesssim \frac{m_\chi}{\rm{GeV}} \lesssim 10^{13}$) does not alter the characteristic angular dependence that we find for the example of $m_\chi = 10^3$ GeV, shown above. The value of the mass will impact the overall amplitude of the signal, which one can appreciate from the overall mass scaling which appears in Eqs.~\eqref{eq:recoilB} and~\eqref{eq:recoilF}, but the angular dependence remains the same.

\section{The Signal of Supersymmetry: ``SUSY Rilles'' from supersymmetric Higher Spins}
\label{sec:SUSY}
Thus far, we have considered scatterings involving separately massive spin-$s$ bosons or
massive spin-$s+1/2$ fermions.
Nevertheless, one can consider models where higher spin particles of both types participate.
Supersymmetry provides a natural framework to study such models, especially at
high energy scales where it is well motivated and becomes a symmetry of the underlying theory. In the context of dark matter, the possibility that a hidden sector dark matter candidate may realize supersymmetry on-shell, i.e., at tree-level in interactions, has recently been studied in  \cite{Burgess:2021juk,Burgess:2021obw}. In particular,  if the dark sector couples to the standard model very weakly, then SUSY breaking in the visible sector does not generate mass splitting in the dark sector at tree level. In this section we adopt this assumption.

Under the assumption of supersymmetry, 
bosonic and fermionic higher spin particles organize in supersymmetric higher spin multiplets.
The Lagrangian description of manifestly supersymmetric, massless, higher spin multiplets 
(see \cite{Gates:2013rka,Buchbinder:2015kca,Buchbinder:2020yip,Koutrolikos:2022chj} and references therein) has been
developed and several types of interactions
\cite{Buchbinder:2017nuc,Koutrolikos:2017qkx,Buchbinder:2018wwg,Buchbinder:2018gle,Buchbinder:2018wzq,Gates:2019cnl} have been constructed. For massive higher spin supermultiplets, which are relevant for our 
discussion, their off-shell description has been  discovered recently \cite{Koutrolikos:2020tel}. The 
off-shell spectrum of such theories is very complicated as it includes towers of auxiliary
superfields with decreasing rank. However on-shell, the only propagating degrees of freedom are
massive bosonic and fermionic higher spin particles like the ones we considered in Sections \ref{hsb}
and \ref{hsf}.

Specifically, there are two types of massive higher spin supermultiplets,
labelled by their superspin value ($\Ysf$), which corresponds to the eigenvalue
of the supersymmetric spin Casimir operator of the $4D, \N=1$ super-Poincar\'{e} algebra.
The first one is the half-integer superspin ($\Ysf=s+1/2$) multiplet which includes a spin-(s+1) boson, 
two spin-$(s+1/2)$ fermions and one spin-$s$ boson. The second one is the integer superspin ($\Ysf=s$) multiplet
with has the following physical degrees of freedom: one spin-$(s+1/2)$ fermion and two spin-$s$ bosons and a 
spin-(s-1/2) fermion. The on-shell spectrum is easily understood from the viewpoint of the supersymmetric algebra. 
For a massive particle, the supersymmetry algebra defines two non-trivial, fermionic,
creation operators 
$Q_1,~Q_2$ which increase spin by $1/2$ units. Therefore acting on a vacuum state with spin-j ($|j\rangle$) 
they generate the states: $Q_1|j\rangle,~Q_2|\rangle,~Q_2Q_1|j\rangle$ which carry spins $j+1/2, j+1/2$ and 
$j+1$. Depending on the spin value of the vacuum state (integer $j=s$ or half-integer $j=s+1/2$) we find the two
supermultiplets mentioned above.

This simple fact, that supersymmetry allows very specific combinations of bosonic and fermionic 
higher spin particles in the spectrum of a theory, will lead to a very characteristic (SUSY Rilles) 
angular dependence and momentum fall-off. Using \eqref{eq:LintBoson} and
\eqref{eq:LintFermion} we get the following interactions for the supersymmetric theory\footnote{Notice that \eqref{eq:LintBoson} includes both spin-$s$ particles of the supermultiplet.}:
\be
\mathcal{L}^{susy}_{int}\sim\mathcal{L}^{(s-1/2)}_{int}+\mathcal{L}^{(s)}_{int}
+\mathcal{L}^{(s+1/2)}_{int}
\ee
Interactions of this type are generated by considering the manifestly supersymmetric interactions between the various superfields that carry the corresponding degrees of freedom. An example is the following superspace action:
\bIea{ll}
\mathcal{S}_{int}\sim\int~d^8z~ \Big\{&g_1~\bar{\Phi}~V~\Phi + 
g_2~\D^{\a_s}\bar{\Psi}^{\a(s-1)\ad(s)}~\Dd^{\bd_s}\Psi^{\b(s)\bd(s-1)}~\pa_{\a_1\ad_1}\dots\pa_{\a_s\ad_s}\pa_{\b_1\bd_1}\dots\pa_{\b_s\bd_s}V\n\label{ssa}\\
&+~g_3~\bar{\Psi}^{\a(s-1)\ad(s)}~\Psi^{\b(s)\bd(s-1)}~\pa_{\a_1\ad_1}\dots\pa_{\a_{s-1}\ad_{s-1}}\pa_{\b_1\bd_1}\dots
\pa_{\b_{s-1}\bd_{s-1}}[\Dd_{\ad_s},\D_{\b_s}]V~\Big\}
\eIea
where $\D_\a, \Dd_{\ad}$ are the superspace spinorial derivatives, $\Phi$ is a chiral superfield that describes the Weyl spinor field $N_\a$ which plays the role of 
matter \big($\D_a\Phi|=N_\a$\big), $V$ is a real scalar superfield that describes the spin-1 mediator
\big(~$[\D_a,\Dd_\ad]V|=G_{\a\ad}$~\big) in two-component spinor notation and $\Psi_{\a(s)\ad(s-1)}$ is a fermionic superfield that carries the higher spin bosons and fermions in two-component spinor notation:
\bIea{l}\n
\frac{1}{(s+1)!~s!}~[\D_{(\a_{s+1}},\Dd_{(\ad_s}]\Psi_{\a(s))\ad(s-1))}|=\psi_{\a(s+1)\ad(s)}~,\sn\\
\Psi_{\a(s)\ad(s-1)}|=\tfrac{1}{M}~\psi_{\a(s)\ad(s-1)}~,\sn\\
\frac{1}{s!}~\Dd_{(\ad_s}\Psi_{\a(s)\ad(s-1))}|=\sigma^{1}_{\a(s)\ad(s)}+i\sigma^{2}_{\a(s)\ad(s)}~.\sn
\eIea
The coupling constants $g_1,~g_2,~g_3$ unify the coupling constants that appear in \eqref{eq:LintBoson} and \eqref{eq:LintFermion}
and the mass parameter $M$ unifies the mass of the higher spin bosons and fermions.

Putting the puzzle pieces together, we arrive at the prediction of higher-spin supersymmetry for directional direct detection with polarized targets. We find,
\be 
\left(\frac{\Delta dR}{dE_Rd\Omega}\right)_{\rm SUSY} = 
\left(\frac{\Delta dR}{dE_Rd\Omega}\right)_s + \left(\frac{\Delta dR}{dE_Rd\Omega}\right)_{s+1/2} +\left(\frac{\Delta dR}{dE_Rd\Omega}\right)_{s-1/2}
\ee 
for the integer superspin ($\Ysf=s$) multiplet, and similarly for the half-integer superspin ($\Ysf=s+1/2$) multiplet. The individual components of the signal are simply those derived in the previous sections, each distinguished by their momentum and angular dependence. This suggests that detector measurements across energy and angles will not only be able to test the spin of a higher-spin dark matter candidate, but also whether they can fit in a supersymmetric multiplet.

Finally, we note that the superspace interaction terms \eqref{ssa} will generate additional interactions besides the ones we 
discussed in Secs.~\ref{hsb} and \ref{hsf}.
These are mixed interactions where the spin-1 mediator interacts with two higher spin particles of different spins. Interesting examples proposed by supersymmetry are the following cubic interactions: (\emph{i}) $(s+1/2) - (s-1/2) - 1$ and (\emph{ii}) $(s+1) - (s-1) - 1$.
\section{Discussion}
\label{sec:discussion}

In this work we have proposed spin-polarized directional direct detection as a probe of the spin of dark matter, generalizing past results of spin-$1/2$ and spin-$1$ to arbitrary spin-$s$ and spin-$(s+1/2)$, and by straightforward extension, to a supersymmetric multiplet of higher-spin fields. From these analyses, we conclude that the polarized double-differential cross section provides a probe the (higher) spin of dark matter through a combination of angular and energy dependence. In summary, for the class of interactions we consider,
\begin{enumerate}
    \item[] {\bf Angular Dependence $\leftrightarrow$ Bosonic vs. Fermionic :} The angular dependence distinguishes integer spin-$s$ and half-integer spin-$(s+1/2)$, independent of $s$.
    \item[] {\bf Recoil Energy Dependence $\leftrightarrow s$:} The bosonic and fermionic cases both exhibit an $s$-dependent fall-off of the rate at low recoil energy as $\frac{d\Delta R}{dE_rd\Omega} \propto E^{2s}$, where $E$ is the recoil energy .
\end{enumerate} 
Based on these features we conclude that the angular and energy dependence of the recoil rate yield complementary information which can be used to determine the spin of massive dark matter particles.

The aim of this work is a modest one, namely to develop the underlying theory of spin-determination applied to higher-spins. We have not touched at all upon the observational prospects for making these measurements. Guided by past work  \cite{Chiang:2012ze,Catena:2018uae} we expect that the exposure times needed for a discovery to be within reach of next generation experiments.  We defer a concrete assessment of experimental prospects and a sensitivity forecast to future work.

In addition to the important work to be done on the detector side, there is much work to be done on the theory side. We have not performed a comprehensive analysis of the set of all possible interactions that can be probed by directional direct detection with spin-polarized targets. An obvious possibility is to take the effective field theory approach and enumerate all possible interactions, as performed in \cite{Dong:2022mcv}, and from this identify those of possible interest. Alternatively, supersymmetry may be taken as a guide, and, as discussed in Sec.~\ref{sec:SUSY} used as a tool to generate new interactions. In particular, one could consider interactions that mix higher-spin fields of differing spin, including the mixing of bosons and fermions, or wherein the spin-$1$ mediator considered here is replaced with a higher spin mediator. We leave the analysis of new interactions to future work. 

An additional possibility is to consider the signal of a {\it tower} of higher-spin fields, with each field acting as a sub-dominant component of the observed dark matter. Given the universality of the angular dependence across varying integer and half-integer spins, searching for a tower of fields will require detailed measurements of the energy dependence of the signal, in order to parse the underlying spectrum of the theory. Under certain assumptions about the mass and spin, e.g., if the fields saturate the Higuchi bound, one might hope for a characteristic signal, or that certain experimental probes become sensitive to particular values of the spin. We leave these interesting possibilities to future work.

Finally, we make a broad comment that this work demonstrates the complementary nature of early universe probes of higher-spin particles, such as the cosmic microwave background, with more traditional particle physics approaches, such as direct detection. This provides a path towards the measurement of the (higher) spin of dark matter.

\vspace{1cm}


{\bf Acknowledgments}\\[.1in] \indent
The authors thank Riccardo Catena, Kare Fridell, Austin Joyce, Wayne Hu, Gordan Krnjaic, Hayden Lee, Wenzer Qin, and Yiming Zhong, for helpful discussions and insightful comments. The work of L.J. is 
supported in part by the Kavli Foundation by a Kavli Fellowship. The work of E.M. is supported in part by 
a Discovery Grant from the National Science and Engineering Research Council of Canada. The work of S.A. 
is supported in part by the Simons Foundation.
S.J.G. was supported in this research in part by the endowment of the Ford Foundation Professorship of Physics at Brown University and the Brown Theoretical Physics Center.  Moreover, he also is grateful to acknowledge the support from the endowment of the Clark Leadership Chair in Science at the University of Maryland - College Park. The work of K.K. is supported in part by the endowment of the Clark Leadership Chair in Science at the University of Maryland, College Park. K.K gratefully acknowledge the hospitality of the Physics Department at the University of Maryland, College Park.
\newpage

\appendix

\pagebreak
\setcounter{equation}{0}
\renewcommand{\theequation}{\thesection.\arabic{equation}}

\section{Higher Spin Polarization Vectors}
\label{spinsPol}
\subsection{Spin-s}
Here we discuss some technical details of the spin-$s$ polarization vectors. A full discussion can be found in \cite{Lee:2016vti} along with further details of higher spin fields in de Sitter space. When we decompose a spin-$s$ field into spatial slices as
\be 
\sigma_{i_1...i_n0...0} = \sum_\lambda \sigma^\lambda_{n,s}\varepsilon^\lambda_{i_1...i_n}, 
\ee 
the polarization tensor, $\varepsilon^{\lambda}_{i_1...i_n}$ must be symmetric, transverse,, and traceless. Using these properties, we can decompose $\varepsilon^\lambda_{i_1...i_n}$ into a transverse and longitudinal component as 
\be 
\epsilon^\lambda_{i_1...i_s}(\hat{\mathbf{k}},\varepsilon) = \varepsilon^\lambda_{(i_1...i_\lambda}(\varepsilon)f_{i_{\lambda+1}...i_s)}(\hat{\mathbf{k}}), 
\ee 
where $\varepsilon^\lambda_{i_1...i_\lambda}$ is the transverse component and $f_{i_{\lambda+1}...i_s}$ is longitudinal. From the transverse component, we can define 
\be 
F^\lambda_s = q_{i_1...i_s}\varepsilon^\lambda_{i_1...i_s}, 
\ee 
such that for three spatial dimensions \be 
F^\lambda_s \propto z \hat{P}^\lambda _s, 
\ee 
where $z = q_{i_1...i_\lambda}\epsilon^\lambda_{i_1...i_\lambda}$ and $P^\lambda_s = \sin^\lambda\theta \hat{P}^\lambda_s$, with $P^\lambda_s$ the associated Legendre polynomials. When $\lambda =0$, we have the property
\be 
q_{i_1...i_s}\epsilon^0_{i_1...i_s}(k) \propto |q|^s P_s(\hat{q}\cdot\hat{k}), 
\ee 
giving rise to the characteristic angular dependence that we discuss. 

Lastly, note that the self contraction of the spin-$s$ polarization tensors is given by 
\be 
\varepsilon^\lambda_{i_1...i_s}\varepsilon^{\lambda*}_{i_1...i_s} = \frac{(2s-1)!!(s+\lambda)!}{2^\lambda ((2\lambda -1)!!)^2 s!(s-\lambda)!}\varepsilon^\lambda_{i_1...i_\lambda}\varepsilon^{\lambda*}_{i_1...i_\lambda}. 
\ee 

\subsection{Spin-s+1/2}
The polarization tensor decomposition for spin-s+1/2 fermions proceeds largely in the same way as for the spin-$s$ bosons as described above. The difference is now that instead of decomposing into $\sigma^\lambda_{n,s}$ and a spin-$s$ polarization tensor, we now decompose into a spin-1/2 mode function and spin-$s$ polarization tensor. We can see this in analogy to the known decomposition for the gravitino. For a spin-3/2 gravitino, $\psi_\mu(\mathbf{p}, \lambda)$ we have \cite{Moroi:1995fs}
\be 
\psi_{\mu}(\mathbf{p}, \lambda) = \sum_{s,m}\Big\langle \left(\frac{1}{2},\frac{s}{2}\right)(1, m)\Big|\left(\frac{3}{2}, \lambda\right)\Big\rangle u(\mathbf{p}, s)\epsilon_\mu(\mathbf{p}, m), 
\ee 
where $\left(\frac{1}{2},\frac{s}{2}\right)(1, m)\Big|\left(\frac{3}{2}, \lambda\right)\Big\rangle$ are the Clebsch-Gordan coefficients, $u$ is a spin-1/2 mode function and $\epsilon_\mu$ is a spin-1 polarization tensor. The spin-s+1/2 case follows directly from the above.

\section{Scattering Amplitude Calculations}
\label{Mcalc}
Here we describe the calculation of the scattering amplitudes for both bosonic and fermionic HSDM-nucleon interactions in further detail. 
\subsection{Bosonic HSDM-Nucleon Interaction}

We consider the following Feynman diagram which follows from our interaction Lagrangian, Eq.~\eqref{eq:LintBoson}. 
\begin{figure}[htb]
     \centering
     \includegraphics[scale=.5]{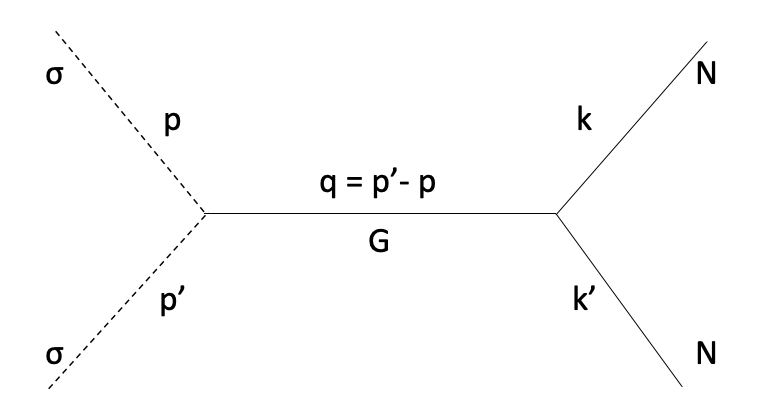}
     \caption{Feynman diagram for HS-nucleon interaction.}
     \label{fig:FeynmanDiagram}
 \end{figure}
Here, $\sigma$ is the HS boson, $N$ is a standard model nucleus and G is the vector mediator. From this diagram, we obtain the following expression for the matrix element, $\mathcal{M}$, 
\be 
i\mathcal{M} = -i\frac{c_s |q|^{2s}}{\Lambda^{2s}m_G^2}P_s(\hat{p}\cdot \hat{q})P_s(\hat{p}'\cdot \hat{q})\left(p^\mu  + p'^\mu \right) \bar{u}_N\gamma_\mu (h_3 + \gamma_5h_4)u_N), 
\ee 
where we have used the fact that the HSDM-mediator vertex is given by 
\be 
V = -i\frac{c_s}{\Lambda^{2s}}\left(p_\nu |p|^{2s} P_s(p\cdot p') + p'_\nu |p'|^{2s} P^*_s(p\cdot p')\right). 
\ee 
This arises from the decomposition of the HS polarization tensors contracted with momenta, discussed in the previous section. Now, we would like to take the non-relativistic limit of this expression. We follow the procedure outlined in \cite{Catena:2018uae} for spin-1 and spin-1/2 DM. First, we take the ingoing and outgoing DM momenta to be:
\begin{align}
    p = m_\chi(1, \Vec{v}), \\
    p' =  m_\chi(1, \Vec{v}'), 
\end{align}
and similarly for the SM nucleus: 
\begin{align}
k = m_N(1, 0),\\
k' = (m_N, -\vec{q}), 
\end{align}
which we take to be initially stationary. The momentum transfer of the system is 
\be 
\vec{q} = m_\chi(\Vec{v}' - \Vec{v}).
\ee 
We also define 
\be 
2 \vec{v}^\perp = \vec{v} + \vec{v}' + \frac{m_\chi}{m_N}(\vec{v}' - \vec{v}). 
\ee 
For the spinor bilinears, one can take the following expansion \cite{}
\begin{align} 
\bar{u}_N(k',r')\gamma_\mu u_N(k,r) &= \begin{pmatrix}
2m_N\delta^{r'r}\\
-\vec{K}\delta^{r'r} - 2i \vec{q}\times \Vec{S}_N^{r'r}
\end{pmatrix},\\
\bar{u}_N(k',r')\gamma_\mu u_N(k,r) &= \begin{pmatrix}
2\vec{K}\cdot \vec{S}_N^{r'r}\\
-4m_N\vec{S}_N^{r'r}
\end{pmatrix}, 
\end{align}
where $\vec{K}$ and $S_N$ are defined respectively as:
\begin{align}
    \vec{K} = \vec{k}+ \vec{k}', \\
    S_N = \xi^{r\dagger}(\vec{\sigma}_N/2)\xi^r.
\end{align}
Putting all of this together, we obtain 
\begin{align}
    -i\mathcal{M} &= -\frac{i}{\Lambda^{2s}}\frac{m_\chi}{m_G^2} \begin{pmatrix} 2 & (\vec{v}' + \vec{v}) \end{pmatrix}\left[h_3\begin{pmatrix}
2m_N\delta^{r'r}\\
-\vec{K}\delta^{r'r} - 2i \vec{q}\times \Vec{S}_N^{r'r}
\end{pmatrix} + h_4\begin{pmatrix}
2\vec{K}\cdot \vec{S}_N^{r'r}\\
-4m_N\vec{S}_N^{r'r}
\end{pmatrix}\right],\\
&=  -\frac{i}{\Lambda^{2s}}\frac{m_\chi m_N}{m_G^2}\delta^{ss'}\left(4h_3 \delta^{rr'} - 8h_4\vec{S}_N^{r'r}\cdot \vec{v}^\perp\right), 
\end{align}
where we have kept terms up to linear order in the momentum. Now, we can sum over the spin states to obtain 
\be 
|\bar{\mathcal{M}}|^2 = \frac{1}{3}\sum_{ss'}\sum_{r'}|\mathcal{M}|^2. 
\ee 
Performing the summation and again keeping terms up to linear order in the momentum, we obtain our final expression for the scattering amplitude:
\be 
|\bar{\mathcal{M}}|^2 = \frac{16m_N^2 m_\chi^2}{m_G^4}\frac{c_s^2}{\Lambda^{4s}} |q|^{4s}P_s(\hat{p}\cdot \hat{q})^2 P_s(\hat{p}'\cdot\hat{q})^2 \left[h_3^2 - h_3h_4 \left(1 - \frac{m_\chi}{m_N}\right)\vec{v}\cdot\vec{s} - h_3h_4\left(1 + \frac{m_\chi}{m_N}\right)\vec{v}'\cdot\vec{s}\right]. 
\ee
\subsection{Fermionic HSDM-Nucleon Interaction}
For the fermionic HSDM we consider the same diagram as Figure \ref{fig:FeynmanDiagram}, where the external dark matter legs are spin-s+1/2 fermions rather than spin-$s$ bosons. Our matrix element is then  
\be 
-i\mathcal{M} \sim -\frac{i}{m_G^2}C_s P_s(\hat{p}\cdot\hat{q})P_s(\hat{p}'\cdot\hat{q}) |q|^{2s} \bar{u}(p)\gamma^\nu \left(\frac{1}{\Lambda_3^{2s}} + \gamma_5\frac{1}{\Lambda_4^{2s}}\right)u(p'), 
\ee 
where we have decomposed our HS fermion via Eq.~\eqref{fermiondecomp}. Then, the spin-1/2 bilinears for the DM will be 
\begin{align}
  i\mathcal{M} &= -\frac{i}{m_G^2}\Bigg[\lambda_3h_3\left(4m_N m_\chi\delta^{s's}\delta^{r'r}\right) + \lambda_3h_4\left(-8m_Nm_\chi\delta^{ss'}\vec{v}^\perp\cdot \vec{S}_N^{r'r} + 8im_N \vec{S}_\chi^{s's}\cdot(\vec{S}_N^{r'r}\times \vec{q})\right)\nonumber \\
  &+\lambda_4h_3\left(-8m_Nm_\chi\delta^{rr'}\vec{v}^\perp\cdot \vec{S}_\chi^{ss'} + 8im_\chi \vec{S}_\chi^{s's}\cdot(\vec{S}_N^{r'r}\times \vec{q})\right) + \lambda_4h_4\left(-16 m_Nm_\chi \vec{S}_\chi^{s's}\cdot \vec{S}_N^{r'r}\right) \Bigg].
\end{align}
Proceeding in the same was as before, we find that the scattering amplitude is given by \begin{align}
  i\mathcal{M} &= -\frac{i}{m_G^2}\Bigg[\Lambda_3h_3\left(4m_N m_\chi\delta^{s's}\delta^{r'r}\right) + \Lambda_3h_4\left(-8m_Nm_\chi\delta^{ss'}\vec{v}^\perp\cdot \vec{S}_N^{r'r} + 8im_N \vec{S}_\chi^{s's}\cdot(\vec{S}_N^{r'r}\times \vec{q})\right)\nonumber \\
  &+\Lambda_4h_3\left(-8m_Nm_\chi\delta^{rr'}\vec{v}^\perp\cdot \vec{S}_\chi^{ss'} + 8im_\chi \vec{S}_\chi^{s's}\cdot(\vec{S}_N^{r'r}\times \vec{q})\right) + \Lambda_4h_4\left(-16 m_Nm_\chi \vec{S}_\chi^{s's}\cdot \vec{S}_N^{r'r}\right) \Bigg].
\end{align}
Then, squaring and summing over the spin states, this reduces to
\begin{align}
    |\bar{\mathcal{M}}|^2 &= \frac{16m_\chi^2m_N^2}{m_G^2}\Bigg\{\Lambda_3^2 h_3^2 + 3\Lambda_4^2h_4^2 \nonumber \\
    &- \left[\Lambda_3^2h_3h_4\left(1- \frac{m_\chi}{m_N}\right) + \Lambda_4^2h_3h_4\left(1+ \frac{m_\chi}{m_N}\right) + 2\Lambda_3\Lambda_4h_4^2\right]\vec{v}\cdot\vec{s}\nonumber \\
    &- \left[\Lambda_3^2h_3h_4\left(1+ \frac{m_\chi}{m_N}\right) + \Lambda_4^2h_3h_4\left(1- \frac{m_\chi}{m_N}\right) - 2\Lambda_3\Lambda_4h_4^2\right]\vec{v}'\cdot\vec{s}\Bigg\}.
\end{align}
\small{
\bibliographystyle{hephys}
\bibliography{HSDM-refs}
}

%
%
%

\end{document}